\documentclass[11pt]{article}
\usepackage{a4wide}
\usepackage[normalem]{ulem}
\usepackage[dvipdfmx]{graphicx}
\usepackage{amsmath,amssymb}
\usepackage{color}
\usepackage{slashed}
\usepackage[bookmarks=true,bookmarksnumbered=true,setpagesize=false]{hyperref}
\usepackage{comment}
\usepackage{cases}

\newcommand{\tx}{\text}
\newcommand{\nn}{\nonumber\\}
\newcommand{\h}{\hspace}
\newcommand{\be}{\begin{equation}}
\newcommand{\e}{\end{equation}}
\newcommand{\aln}[1]{\begin{align}#1\end{align}}
\newcommand{\paren}[1]{\left(#1\right)}

\newcommand{\br}[1]{\left\{#1\right\}}

\newcommand{\wt}{\widetilde}
\newcommand{\hyphen}{\mathchar`-}

\begin{document}
\title{
\vspace{-2cm}
\vbox{
\baselineskip 14pt
\hfill \hbox{\normalsize KEK-TH-2317
}} 
\vskip 1cm
\bf \Large Axion-CMB Scenario in Supercooled Universe
%QCD Axion as an origin of CMB anisotropy in $B-L$ Model with Symmetry Non-Restoration 
\vskip 0.5cm
}
\author{
 Satoshi Iso$^{ab}$\thanks{E-mail: \tt iso(at)post.kek.jp},\h{1mm} 
 Kiyoharu~Kawana$^{ac}$\thanks{E-mail: \tt kawana(at)snu.ac.kr}\h{1mm} and 
 Kengo Shimada$^{a}$\thanks{E-mail: \tt kengo.shimada(at)kek.jp}
\bigskip\\
\it 
\normalsize
 $^a$ Theory Center, High Energy Accelerator Research Organization (KEK),\\
 \normalsize 
\it  $^b$ Graduate University for Advanced Studies (SOKENDAI),\\
 \normalsize
\it Tsukuba, Ibaraki 305-0801, Japan, \\
\normalsize 
\it  $^c$ Center for Theoretical Physics, Department of Physics and Astronomy,\\
 \normalsize
\it  Seoul National University, Seoul 08826, Korea,
\smallskip
}

\date{\today}

\maketitle

\begin{abstract}\noindent
"{\it Axion-CMB scenario}" is an interesting possibility to explain the temperature anisotropy of 
the cosmic microwave background (CMB)  by primordial fluctuations of the QCD axion \cite{Iso:2020pzv}.   
In this scenario, fluctuations of radiations are generated by an energy exchange between axions and radiations, which results in the correlation  between the primordial axion fluctuations  and the CMB anisotropies.  
Consequently, the cosmological observations stringently constrain a model of the axion and the early history of the universe.  
In particular, we need a large energy fraction $\Omega_A^{}$ of the axion at the QCD phase transition, but it must become tiny at the present universe to suppress the isocurvature power spectrum.  
 One of natural cosmological scenarios to realize such a situation
  is the thermal inflation which can sufficiently dilute the axion abundance. 
 Thermal inflation occurs in various models.
 In this paper, we focus on 
a classically conformal (CC) $B$-$L$ model with a QCD axion.
In this model, 
the early universe undergoes a long supercooling era of the $B$-$L$ and electroweak  symmetries, and 
thermal inflation naturally occurs. 
Thus it can be a good candidate for the axion-CMB scenario.
But the axion abundance at the QCD transition is shown to be insufficient in the original  CC $B$-$L$ model.
To overcome the situation, we extend the model by introducing $N$ scalar fields $S$ (either massive or massless)
and consider a novel cosmological history such that the $O(N)$  and the $B$-$L$ sectors evolve almost separately in the early universe. 
We find that all the necessary conditions for the axion-CMB scenario can be satisfied 
in some parameter regions for massless $S$ fields, typically
{$N\sim 10^{19}$} and the mass of $B$-$L$ gauge boson around $5-10$ TeV. 
\end{abstract}

\newpage

%\normalsize
%________________________________________
\section{Introduction}
Anisotropy of the cosmic microwave background (CMB) is one of the most fascinating subject
 in the particle  cosmology since it is generated at an early stage of the history of the unvierse
 and possibly related to the physics beyond the Standard Model (BSM).  
The current observational data such as Planck 2018 \cite{Aghanim:2018eyx,Akrami:2018odb,Akrami:2019izv} tells us that the temperature fluctuation is almost scale invariant and adiabatic, which favors inflation models by a single scalar field. 
But it is  is not the unique  scenario for explaining the CMB anisotropy.  
A well known example is the curvaton scenario 
\cite{Lyth:2001nq,Enqvist:2001zp,Moroi:2001ct,Lyth:2002my,Gordon:2002gv,Dimopoulos:2003az,Sasaki:2006kq,Chingangbam:2009xi,Enqvist:2009ww,Nakayama:2009ce,Byrnes:2010xd} where the origin of fluctuations comes from an 
additional scalar field called curvaton.   
In this scenario, the primordial fluctuation of curvaton is converted to that of radiation through decay of curvatons.  
An interesting aspect of the curvaton model is a prediction of sizable non-Gaussianities.
% even if the primordial fluctuation is Gaussian due to  the non-trivial dynamics of the conversions. 
%

Axion-CMB scenario is  similar to the curvaton scenario where 
% we recently studied a  possibility 
CMB anisotropy is induced by the primordial perturbations of the QCD axion   \cite{Iso:2020pzv}.
In this scenario, Peccei-Quinn symmetry \cite{Peccei:1977hh,Wilczek:1977pj,Weinberg:1977ma,Kim:1979if,Shifman:1979if,Zhitnitsky:1980tq,Dine:1981rt,Kim:2008hd,diCortona:2015ldu} is assumed to be already broken before the primordial inflation, 
and the axion field acquires primordial fluctuations during the inflation.  
As the universe cools down to the QCD scale $T_{\rm QCD}^{}$, the axion  potential is generated
by transferring non-zero energy from radiation. Then
 the primordial axion  fluctuations are  converted to the  density  fluctuation of the radiation.  
For a successful realization of the scenario,    three conditions must be satisfied:
(1) A large amount of energy density of axions at $T=T_{\rm QCD}^{}$ is necessary to suppress non-Gaussianity. 
(2)  Axions must be largely diluted until present in order to satisfy the isocurvature constraint.  
(3) In addition, it must explain the observed CMB amplitude.
These three conditions require something like  thermal inflation \cite{Lyth:1995ka,Gong:2016yyb,Hambye:2018qjv,Baratella:2018pxi} 
after the QCD temperature \cite{Iso:2020pzv}.
%so that axion abundance is sufficiently diluted     
%\cite{Lyth:1995ka}-\cite{Baratella:2018pxi} 
%
Indeed, under a couple of reasonable assumptions, we found that they can be satisfied as long as the thermal inflation
 lasts long enough after the QCD phase transition. 
The  purpose of this paper is to present a concrete  particle physics model  that can realize such a thermal inflation.  
In this paper, we will consider the classically conformal $B$-$L$ model  
\cite{Iso:2009ss,Iso:2009nw,Iso:2012jn,Iso:2017uuu} with a QCD axion and its extension with $O(N)$ scalar fields $S$. 
Originally, the $B$-$L$ model with classical conformality was proposed to explain the eletroweak (EW) scale by Coleman-Weinberg (CW) mechanism \cite{Iso:2009ss,Iso:2009nw,Iso:2012jn,Hamada:2020wjh,Hamada:2021jls}. 
Besides, it was  pointed out that this model  predicts a long supercooling era of the $B$-$L$ and EW symmetries due to the classical conformality of the scalar potential \cite{Iso:2017uuu}. 
The supercooling  lasts  below $T_{\rm QCD}^{}$ and the energy density of radiation becomes 
smaller than the vacuum energy of the false vacuum. Thus thermal inflation occurs even after the QCD phase transition
and axions can be sufficiently diluted after the QCD phase transition. 
But there is one technical difficulty to obtain a  large value of the axion abundance at $T=T_{\rm QCD}^{}$.
It is because the Higgs vacuum expectation value (vev), $\langle h\rangle|_{T_{\rm QCD}}  \sim \Lambda_{\rm QCD}$, which
is responsible for the nonzero axion potential,  is generated by quark condensates $\langle \bar{q} q \rangle$ 
and becomes of order $100$ MeV. 
As we will see, it is too small for the necessary value of the axion abundance. 
In order to overcome this difficulty, we need an additional mechanism to raise
 $\langle h\rangle|_{T_{\rm QCD}}$ to at least $100$ GeV.  

For this purpose, we introduce  $O(N)$ scalar fields $S$  
coupled to the Higgs field with a very weak negative coupling, 
and consider a novel history of the early universe:
we suppose that  the $O(N)$ sector has evolved almost separately from our universe, 
with a much higher temperature $ \tilde{T}$ than that of our universe. 
Then the  negative coupling will generate sizable negative thermal corrections to the Higgs quadratic potential \cite{Weinberg:1974hy,Meade:2018saz,Baldes:2018nel,Glioti:2018roy,DAgnolo:2019cio,DiLuzio:2019wsw,Matsedonskyi:2020mlz,Chaudhuri:2020xxb}, and $\langle h\rangle|_{T_{\rm QCD}}$ can become as large as  $100$~GeV. 
In the paper, we will consider two extremal cases, very massive $S$ with $m_S^{} \gg m_{Z^\prime}$ or massless $S$. 
We show that various observational constraints, especially sufficient dilution of axions and $S$ particles, 
 can be simultaneously satisfied in the massless case for {$N\gtrsim 10^{19}$}, but not in the massive case.  
One might think that such a large number of degrees of freedom (dof) would be inconsistent with   various phenomenological aspects such as collider observables. 
 This is actually problematic as long as we focus on moderate values of the portal coupling $\lambda_{SH}^{}$. 
 However, as discussed in Refs.~\cite{Meade:2018saz,Glioti:2018roy}, it is possible to make a model consistent by taking $\tilde{\lambda}_{SH}^{}:=N\lambda_{SH}^{}$ as a free parameter and choosing moderate values of $\tilde{\lambda}_{SH}^{}$.~ 
 Roughly speaking, collider observables are determined by the combination $N\lambda_{SH}^{2}$ and this is suppressed by $1/N$ when we fix $\tilde{\lambda}_{SH}^{}$.  
 See Refs.~\cite{Meade:2018saz,Glioti:2018roy} for more  detailed discussion.

The organization of the paper is as follows.  
In section \ref{sec:review}, we review the axion-CMB scenario \cite{Iso:2020pzv} and explain what conditions are necessary for the scenario to be observationally viable. 
In section \ref{sec:PPmodels}, we focus on a specific particle physics model of the axion-CMB scenario,
 a classically conformal (CC)  $B$-$L$ model since the model predicts 
 thermal inflation below $T_{\rm QCD}^{}$. 
We first see that  the  original CC $B$-$L$ model has a problem as a candidate for  the axion-CMB scenario. 
We then propose an extended model to overcome this difficulty by introducing an additional $O(N)$ scalar field $S$,  
and investigate the model for either large $m_S^{} \gg m_{Z^\prime}$ or $m_S^{}=0$. 
Our analysis shows that the massless case can satisfy all the necessary conditions for the model to be phenomenologically viable.

%%%%%%%%%%%%%%%%%%%%%%%%%%%%%%%%%%%%%%%%%%%
%________________________________________
\section{Axion-CMB Scenario}\label{sec:review}
In this section, we summarize the axion-CMB  scenario \cite{Iso:2020pzv}. 
It is similar to the curvaton scenario, but the  transfer mechanisms of the fluctuations from 
the curvaton (or axion) to radiation are different. Curvatons decay into radiation to generate the CMB
anisotropy while axions are assumed to be stable until present. 
In the axion-CMB scenario, the fluctuations of radiation are induced when the axion potential is generated at the QCD phase transition. 
Such scenario is discussed in Ref.~\cite{Hertzberg:2008wr} but it is shown to be inconsistent with observations as far as the standard cosmology is concerned.  
The reason is the following. 
For the scenario to be observationally viable, the model must satisfy the following
three conditions:
\begin{enumerate}
\item production of  sufficient amplitudes of the CMB anisotoropy
\item   consistency with the axion isocurvature constraint
\item   consistency with the non-Gaussianity constraint. 
\end{enumerate}
The conditions, 1 and  3, require that a sufficiently large energy fraction $\Omega_A$ of the axion is 
present at the QCD scale when the axion-potential is generated.
On the other hand,  the axion abundance must be tiny at the present era to satisfy the isocourvature constraint. 
In order to satisfy them simultaneously, we need a mechanism to dilute axions after the QCD phase transition
such as low scale thermal inflation. 

The axion-CMB scenario is effectively parametrized by three parameters:
\begin{itemize}
\item Amplitude of the primordial fluctuations of axion
     $(\delta A/\bar{A})^2 \sim H^2_{\text{exit}}/(f_A \theta)^2$, 
\item ratio of energy densities of axion to that of radiation right after the axion potential is generated, 
$R=\Omega_A/\Omega_r$, 
\item fraction of the  axion abundance in the total cold dark matter (CDM) today,
 $r_A=\Omega_A/\Omega_{\rm CDM}|_{\text{today}}$. 
 \end{itemize}
The three conditions constrain the allowed region of the three three parameters, together with the initial misalignment angle $\theta_{\rm ini}$. 
We first summarize them in the following. 
See Ref.~\cite{Iso:2020pzv} for more details.

\vspace{5mm}
%%%%%%%%%%%%%%%%%%%%%%%%%%%%%%%%%%%%%%
\noindent$\bullet$ {\bf CMB amplitude}\\
QCD-like axion  $A$ is assumed to be massless during the primordial inflation and fluctuates with the amplitude $\delta A_{\rm ini}$; 
\aln{
\langle \delta A_{{\rm ini}}(k) \delta A_{{\rm ini}}(k') \rangle = (2\pi)^{3} \delta^{(3)}(k+k') \frac{H^{2}_{\rm exit}(k)}{2 k^{3}} ~,~~~~ H_\text{exit}^{}(k) := H|_{k=a\overline{H}}^{} ~.
\label{gaussianA}
}
The axion  acquires potential  $V_A^{}(A)$ at the QCD temperature % \red{$T=T_A^{}$ }
and becomes massive. 
In the generation of the axion potential, increase of the potential energy is compensated by decrease of radiation energy. 
Thus the primordial axion fluctuations induce the density fluctuations of radiations.  
Suppose that the dominant part of the density fluctuations  {of radiation} originate in this induced fluctuations and their initial curvature perturbation is negligible.   
Then the curvature perturbation is given by the axion fluctuation as 
\be
 \zeta_{r}^{}  %= \zeta_{\rm TI} - \frac{1}{4} \ln \left( 1 -  R^{} \frac{\rho_{A} - \bar{\rho}_{A} }{\bar{\rho}_{A}}\right)
  \sim \frac{R}{4}\frac{\delta V_A^{}}{V_{\bar A}}  \delta A_{\rm ini}^{}~,
\label{adiabatic-mode}
\e
where $V_{\bar A}= V_A (A={\bar A})$ and 
\be
%R^{} := \frac{ \Omega_{A}^{}}{ \Omega_{r}^{}} \bigg|_{\rm right~after~transition}^{} 
R^{} := \frac{ \rho_{A}^{}}{ \rho_{r}^{}} \bigg|_{T_{\rm QCD}}^{} 
\label{def of R}
\e
is the ratio of the energy densities of the axion to radiation evaluated right after the  potential is 
instantaneously generated at $T=T_{\rm QCD}^{}$. 
If $\rho_A^{}$ is dominated by the potential energy of QCD axion  at this moment,  $R$ can be calculated as~\cite{Iso:2020pzv} 
\aln{
R& %=\frac{\rho_A^{}}{\rho_r^{}}\bigg|_{T=T_{\rm QCD}^{}}
=\frac{V_A^{}(\overline{A})}{\rho_r^{}}\bigg|_{T=T_{\rm QCD}^{}}\simeq  \frac{30}{\pi^2 g_{\rm QCD}^{}}\frac{m_u^{}/m_d^{}}{(1+m_u^{}/m_d^{})^2}\frac{m_\pi^2f_\pi^2}{T_{\rm QCD}^4}  
(1-\cos (\overline{\theta}_{\rm ini}) )\nn \\
&\simeq 0.012\times \left(\frac{150{\rm MeV}}{T_{\rm QCD}^{}}\right)^4  \left(\frac{\langle h\rangle }{246~{\rm GeV}}\right)
(1-\cos (\overline{\theta}_{\rm ini})) .  
\label{typical R}
}
where $g_{\rm QCD}^{}=69/4$ is the effective number of dof right after the QCD phase transition to which pions also contribute. 
Note that the Higgs vev is treated 
as a free parameter in the above equation since 
this point becomes the most crucial in discussing an explicit
 realization of the scenario in a concrete particle physics model discussed in the next section.    

{The above curvature perturbation  $\zeta_r^{}$ is passed down to the current density of radiation} 
if there are no further mixings with other fields, and 
the CMB amplitude  is given by 
\aln{
\sqrt{A_s} = %\sqrt{{\cal{P}}_{{\cal RR}}^{}(k_*^{})}=
R   \frac{H_\text{exit}^{}(k_*^{})}{4\pi f_A \overline{\theta}_{\rm ini}} 
%{\cal X}(\bar{\theta}_{\rm QCD}) {\cal Y}(\bar{\theta}_{\rm QCD})  ~.
%
=\sqrt{2.1\times 10^{-9}}=4.6\times 10^{-5}~,
\label{CMB by axion}
}
where $k_*^{}=0.05\text{Mpc}^{-1}$ is the reference (pivot) scale. 
This gives a relation between $R$ and $H_\text{exit}^{}(k_*^{}) / \pi f_A$ for each misalignment angle $\bar{\theta}_{\rm ini}^{}$. 

\

%%%%%%%%%%%%%%%%%%%%%%%%%%%%%%%%%%%%%%
\noindent$\bullet$ {\bf Non-Gaussianity}\\
The anharmonicity of the axion potential  produces non-Gaussianities of the CMB anisotropy.
Following Ref.~\cite{Iso:2020pzv}, we have 
\aln{
&f_{\rm NL}^{}\sim -\frac{10}{3R}\frac{\cos (\overline{\theta}_{\rm ini}) \left( 1-\cos (\overline{\theta}_{\rm ini}) \right)}{\sin^2 (\overline{\theta}_{\rm ini} ) } -\frac{10}{3} ~,
\\
&g_{\rm NL}^{}\sim -\frac{1}{6}\left(\frac{20}{3R}\right)^2%\frac{(1-\cos\theta)^2}{\sin^2\theta}
\tan^2\left(\frac{\overline{\theta}_{\rm ini}} {2}\right)+\frac{20}{3}f_{\rm NL}^{}-\frac{1}{6}\left(\frac{20}{3}\right)^2~.
} 
The non-Gaussianities are inversely proportional to $R$, beucuse 
 both of the leading Gaussian fluctuation and sub-leading nonlinear parts are
 proportional to $R\ (<1)$.  Consequently  non-Gaussiannities, by definition,  become tiny for a small value of $R$. 
Thus the observational non-Gaussianity constraints favor large $R$ region.

%%%%%%%%%%%%%%%%%%%%%%%%%%%%%%%%%%%%%%%%%
\begin{figure}[t!]
\begin{center}
\includegraphics[width=8cm]{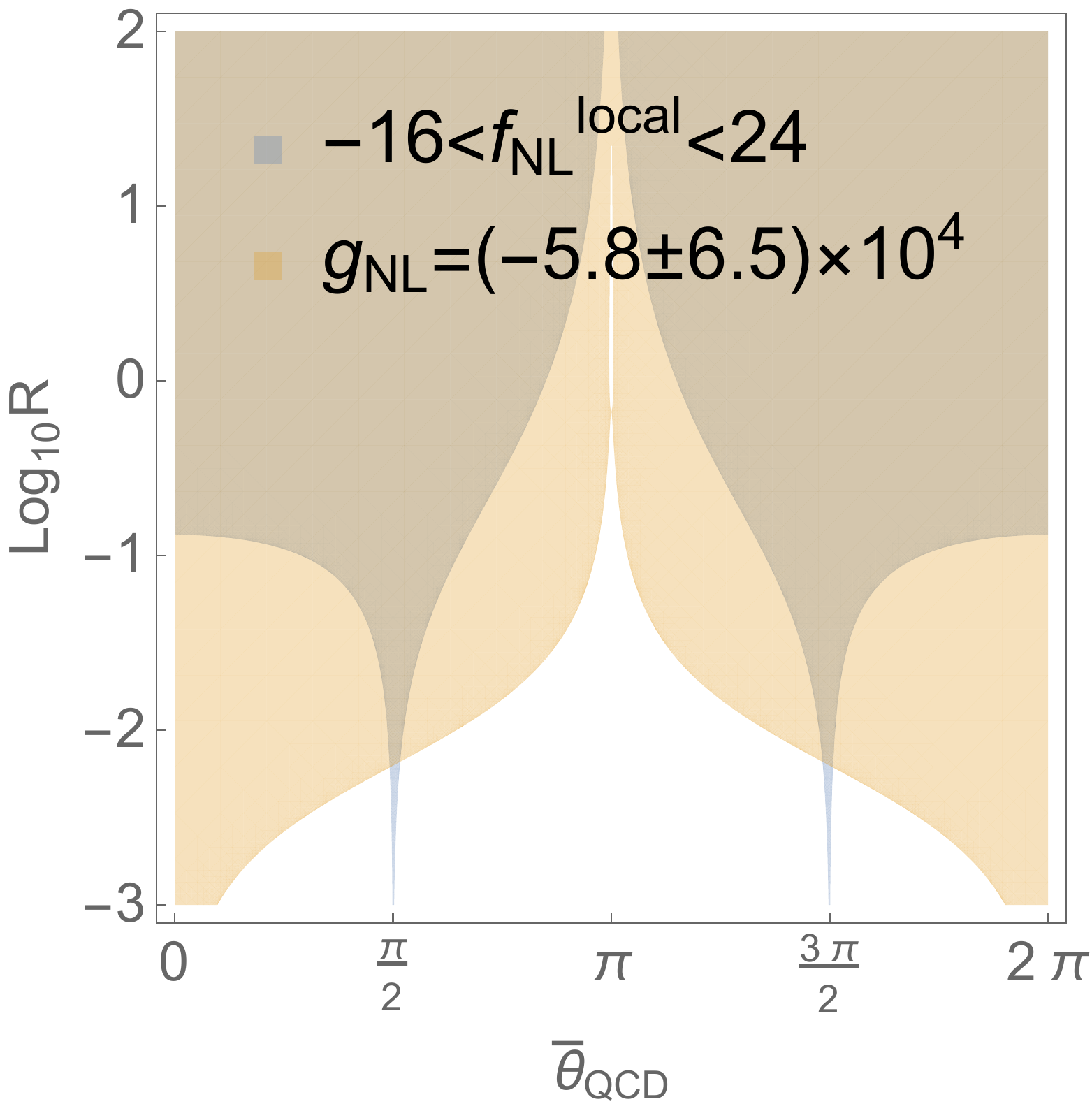}
\caption{
Allowed regions of $(\overline{\theta}_{\rm QCD},R )$ by the non-Gaussianity constraints, Eq.(\ref{f and g}). 
The blue (orange) region represents the region allowed by 
$f_{\rm NL}^{}\ (g_{\rm NL}^{})$. 
$\overline{\theta}\sim \pi/2,3\pi/2$ is necessary to obtain $R \sim {\cal{O}}(0.01)$, where $V''$ vanishes.  
}
\label{fig:NG}
\end{center}
\end{figure}
%%%%%%%%%%%%%%%%%%%%%%%%%%%%%%
In Fig.~\ref{fig:NG}, we plot the allowed region of $(\overline{\theta}_{\rm ini}, R)$ determined by the observational 
 bounds for the non-Gaussianity \cite{Akrami:2019izv} 
\aln{
f_{\text{NL}}^{\rm local}=4\pm 20,%-0.9\pm 10 %5.1 
\quad g_{\rm NL}^{}= (-5.8\pm 13 %6.5
)\times 10^4,\quad (95\%\text{CL by Planck 2018}).
\label{f and g}
}
For more details, see Ref.~\cite{Iso:2020pzv}.  % in Eq.~(\ref{f and g}).  
The blue (orange) region corresponds to $f_{\rm NL}^{}\ (g_{\rm NL}^{})$.  
One can see that, as long as $\overline{\theta}_{\rm ini} \sim 0$, the lower bound of $R$ is ${\cal{O}}(0.1)$, but the bound can be reduced to $\ {\cal{O}}(0.01)$ around $\overline{\theta}_{\rm ini}=\pi/2$ or $3\pi/2$. 
From Eq.~(\ref{typical R}), these lower bounds of $R$ requires the Higgs vev $\langle h \rangle$ at the QCD phase transition to be larger than ${\cal O}(1)$ TeV for $\theta_{\rm ini} \sim 0$, or larger than ${\cal O}(10^2)$ GeV for $\theta_{\rm ini} \sim \pi/2$ or $3 \pi/2$.  
Using Eq.~(\ref{CMB by axion}), the bound $R\gtrsim 0.01$ corresponds to 
\aln{
%{\cal X}(\overline{\theta}_{\rm QCD}) {\cal Y}(\overline{\theta}_{\rm QCD})  
\frac{H_\text{exit}^{}(k_*^{})}{4\pi f_A^{} \overline{\theta}_{\rm ini}} < 4.6 \times 10^{-3}~.
}

\

%%%%%%%%%%%%%%%%%%%%%%%%%%%%%%%%%%%%%
\noindent$\bullet$ {\bf Isocurvature perturbations}\\
The fluctuation of the axion field also produces isocurvature perturbations as in the standard cosmology of QCD-axion \cite{Preskill:1982cy,Abbott:1982af,Dine:1982ah,Beltran:2006sq,Hertzberg:2008wr,Wantz:2009it,Hikage:2012be,Kawasaki:2013ae}.  
The isocurvature power spectrum is calculated as 
\aln{
{\cal{P}}_{{\cal II}}^{}(k)=\frac{k^3}{2\pi^2}P_{{\cal{I}}}^{}(k)=
\left( %{\cal Y}(\bar{\theta}_{\rm QCD}) {\cal Z}(\bar{\theta}_{\rm QCD})~
\frac{r_A^{}H_{\rm exit}(k)}{\pi f_A \overline{\theta}_{\rm ini}}\right)^2 ~,
}
where $r_A^{}$ is the ratio of the abundance of the axion to the total cold dark matter (CDM) today,
\aln{
r_A^{}:=\frac{\Omega_A^{}}{\Omega_{\rm CDM}^{}}\bigg|_{\text{today}}^{},
%\quad \Omega_{\rm CDM}^{}=\Omega_d^{}+\Omega_A^{}~,
}
%\red{where $\Omega_d^{}$ denotes the energy fraction of unspecified component of CDM. }
where $\Omega_{\rm CDM}^{}$ denotes the energy fraction of  CDM. 
By plugging this into 
the isocurvature constraint  \cite{Akrami:2018odb},  
\aln{
& \beta_{\rm iso}(k) = \frac{{\cal P}_{\cal II}}{{\cal P}_{\cal RR} +  {\cal P}_{\cal II}}   < 0.00107 ~~~{\rm for}~ \cos \Delta = -1~,
\nonumber \\
& \cos \Delta = \frac{{\cal P}_{\cal RI}}{\sqrt{{\cal P}_{\cal RR} {\cal P}_{\cal II}}}~,
}
%By plugging this into Eq.~(\ref{correlated}) 
and using ${\cal{P}}_{{\cal RR}}^{}(k_*^{})=2.1\times 10^{-9}$, we obtain the following constraint 
\aln{
%{\cal Y}(\bar{\theta}_{\rm QCD}) {\cal Z}(\bar{\theta}_{\rm QCD})~ 
\frac{r_A^{}H_{\rm exit}^{}(k_*^{})}{\pi f_A^{} \overline{\theta}_{\rm ini}}<1.5\times 10^{-6}
\quad \text{ for } \cos\Delta=-1~.   
\label{isocurvature constraint}
}
In particular, by eliminating
$H_\text{exit}^{}(k_*^{})/(f_A^{}\overline{\theta}_{\rm ini})$ from Eqs.~(\ref{CMB by axion}) and (\ref{isocurvature constraint}),  we obtain an inequality between $r_A^{}$ and $R$ as 
\aln{
r_A^{}<
8.2\times 10^{-3}\ R %\frac{{\cal X} (\overline{\theta}_{\rm QCD})}{{\cal Z}(\overline{\theta}_{\rm QCD})}
~,
\label{r and R}
}
which shows that the axion abundance has to be sufficiently diluted after the QCD transition.  
Thus we need a mechanism such as thermal inflation at the QCD scale. % with a dilution factor $\lesssim {\cal O}(10^{-2})$.

In the previous paper \cite{Iso:2020pzv}, we have studied thermal inflation scenario 
 and shown that all the necessary conditions can be satisfied under reasonable assumptions.  
Thus, a next step is to construct a concrete model of particle physics. 
In the next section, we will consider the classically conformal $B$-$L$ model  
\cite{Iso:2009ss,Iso:2009nw,Iso:2012jn,Iso:2017uuu} with a QCD axion
because the model  undergoes supercooling of the $B$-$L$ and EW symmetries, 
and low scale thermal inflation naturally occurs. 
However, as we will see in the next section, the model predicts a small Higgs vev $\langle h \rangle|_{T_{\rm QCD}} \sim \Lambda_{\rm QCD}$ when the axion potential is generated. 
Consequently  the height of 
the axion potential is  too low to get a sufficiently large value of $R\sim 0.01$, and
the model is already excluded from  the non-Gaussianity constraint.   
Hence, we need some modifications of the original  $B$-$L$ model so that $\langle h \rangle|_{T_{\rm QCD}}$ becomes
at least ${\cal O}(10^2)$ GeV, or larger.

%%%%%%%%%%%%%%%%%%%%%%%%%%%%%%%%%%%%%%%%
%________________________________________
\section{Particle Physics Models of Axion-CMB Scenario}
\label{sec:PPmodels}
As we saw in the previous section, we need a larger value of $R$ than $\sim 0.01$, which corresponds to  $\langle h \rangle|_{T_{\rm QCD}} \sim {\cal O}(10^2)$ GeV. 
In addition, a dilution mechanism like thermal inflation is necessary to satisfy a small value of $r_A^{}$; 
 $r_A^{}< 8.2 \times 10^{-3} R$. 
 In this section, we consider a classically conformal $B$-$L$ model
since the low scale thermal inflation naturally occurs.

%%%%%%%%%%%%%%%%%%%%%%%%%%%%%%%%%%%%%%%%%%%%%%%%%%%%%%%%
\subsection{Classically conformal $B$-$L$ model}
The classically conformal (CC) $B$-$L$ model \cite{Iso:2009ss,Iso:2009nw,Iso:2012jn,Iso:2017uuu} is an extension of the SM with
the right handed neutrinos $N_i^{}$, the $B$-$L$ gauge boson $Z^\prime$, and the $B$-$L$ scalar $\Phi$ which breaks the $B$-$L$ gauge symmetry by Coleman-Weinberg (CW)  mechanism.     
Real components of scalar fields are represented as $h$ and $\phi$ respectively.
We first summarize the thermal history of the early universe of the model.  

For this purpose we need the behavior of the scalar potential. 
At zero-temperature, the effective potential  is given,  at one-loop level, by \cite{Iso:2017uuu}
\aln{
V=V_{\rm TI}^{}+\lambda_H^{}(H^\dagger H)^2-\lambda_{\phi H}^{}(H^\dagger H)(\Phi^\dagger \Phi)
+V_{\text{CW}}^{}(\Phi, H) ~.
}
The SM and $B$-$L$ sectors couple through the $U(1)$ gauge  and the scalar mixings. 
We will focus on the CW potential for  the $\phi$ field by setting $h=0$, 
\aln{
V_{\rm CW}^{}(\phi)=&\frac{B}{32\pi^2}\phi^4\ln \left(\frac{\phi}{v_\phi^{}e^{1/4}}\right),
\quad 
B\simeq  3(2g_{B-L}^{})^4-2\text{Tr}[(\hat{Y}_N^{}/\sqrt{2})^4]~.  \label{CWpotential}
}
Here, $g_{B-L}^{}$ is the $B$-$L$ gauge coupling, $Y_{N,ij}$ is the yukawa coupling between $\Phi$ and $N_i^{}$, 
and $v_\phi^{}$ represents the minimum of the potential $V_{\rm CW}(\phi)$.   
The vev will be shifted by the scalar mixing in a nonzero Higgs vev, 
but its effect to $V(\phi)$ is tiny for $\langle h \rangle \ll \langle \phi \rangle$
and  neglected in the following discussions. 
The parameter $V_{\rm TI}^{}$ is  chosen so that the total vacuum energy vanishes:   
\aln{
V_{\rm TI}^{}=\frac{B}{128\pi^2}v_\phi^4~,
\label{vacuum energy}
}
which determines the Hubble scale of thermal inflation discussed below.
The $B$-$L$ symmetry breaking triggers the EW symmetry breaking via the scalar mixing with a negative coefficient.
Since the Higgs potential becomes
\aln{
\frac{\lambda_H^{}}{4}&\left(h^4-\frac{\lambda_{\phi H}^{}}{\lambda_H^{}}v_\phi^2h^2\right)=\frac{\lambda_H^{}}{4}\left(h^2-\frac{\lambda_{\phi H}^{}}{2\lambda_H^{}}v_\phi^2\right)^2-\frac{\lambda_{\phi H}^2}{4\lambda_H^{}}v_\phi^4~,
}
the Higgs vev is given by
\aln{
&\ v_{h}^{}=\sqrt{\frac{\lambda_{\phi H}^{}}{2\lambda_H^{}}}\times v_\phi^{}=246~\text{GeV}~,
\label{Higgs vev}
}
which gives a relation  between $\lambda_{\phi H}^{}$ and $v_\phi^{}$.  

Next, let us consider its thermal effects. 
Including the thermal correction of $Z^\prime$ and $N_i^{}$, the high temperature expansion of one-loop effective potential of $\phi$ is given by \cite{Iso:2017uuu}
\aln{
V_{\rm 1loop}^{}(\phi)&=\frac{c_2^{}}{2}T^2\phi^2-\frac{c_3^{}}{3}T\phi^3+\frac{\tilde{B}}{4}\phi^4~,
%\frac{T^2M_{Z^\prime}(\phi)^2}{8}-\frac{TM_{Z^\prime}(\phi)^3}{4\pi}+\frac{3M_{Z^\prime}^{}(\phi)^4}{32\pi^2}\ln\left(\frac{T}{m_{Z^\prime}^{}c}\right)~,
}
where
\aln{
&c_2^{}=g_{B-L}^2+\frac{1}{24}\text{Tr}(\hat{Y}_N^{\dagger}\hat{Y}_N^{}),\ \ \  c_3^{}=\frac{6}{\pi}g_{B-L}^3,\ 
\nonumber \\
& \tilde{B}=\frac{1}{8\pi^2}\left\{3(2g_{B-L}^{})^4\ln\left(\frac{T}{m_{Z^\prime}^{}c}\right)-2\text{Tr}\left[(\hat{Y}/\sqrt{2})^4\ln\left(\frac{T}{m_{N}^{}c}\right)\right]\right\}~,
\label{one loop of phi}
}
and 
\aln{
&m_{Z^\prime} =2g_{B-L}^{} v_\phi^{},  \ \ 
m_{N}^{}=Y_N^{}v_\phi^{}/\sqrt{2},
 \nonumber \\
& c=(e\alpha_B^{})^{-1/2},\  \ 
 \log \alpha_B^{-1/2}=-\log 4\pi+\gamma_E^{}. 
 \label{mzmn}
}
In the following, we assume $g_{B-L}^{}\gg Y_N^{}$ so that we can neglect the contributions from $N_i^{}$ for simplicity. 
At high temperature, the potential has a minimum at $\phi=h=0$. 
When $T$ decreases, a new minimum appears and,  below $T_c^{}=ce^{4/3}m_{Z^\prime}^{}$,  the potential height at the new minimum becomes lower. 
In usual cases, the false minimum at $\phi=0$ becomes unstable around this temperature. 
But in the class of models with classical conformality, 
there are no quadratic terms in the zero-temperature scalar potential, 
and  the coefficient of the quadratic term of finite-temperature potential is always positive,
 and hence the false vacuum $\phi=0$ remains the local minimum at any smaller temperature below $T_c$;
the early universe experiences a long supercooling era of the would-be broken symmetries. 
Finally quantum tunneling or some other effects destabilize the false vacuum. 

During the supercooling era, temperature of the universe drops  preserving $B$-$L$ and EW symmetries,  and
when the vacuum energy in Eq.~(\ref{vacuum energy}) dominates the radiation energy, 
%In particular, when $g_{B-L}^{}\lesssim 0.2$, this situation lasts for a long period of time, 
thermal inflation begins. 
By using Eqs.~(\ref{CWpotential}) and (\ref{vacuum energy}),  
the temperature when the thermal inflation starts is given by
\aln{
%V_{\rm TI}^{}\sim \frac{\pi^2 g_{\rm TI}^{} T_\text{inf}^4}{30}\quad \therefore
 T_\text{TI}^{}=\left(\frac{30 V_{\rm TI}^{}}{\pi^2 g_{\rm TI}^{}}\right)^{1/4}= \left(\frac{45}{64 g_{\rm TI}^{}}\right)^{1/4}\frac{m_{Z^\prime}^{}}{\pi}~,%=\left(\frac{90}{128g_{\rm TI}^{}}\right)^{1/4}\frac{m_{Z^\prime}^{}}{\pi}
}
where $g_{\rm TI}^{}$ is the degrees of freedom (dof) at $T=T_{\rm TI}^{}$. 
%
%Without a fine-tuning of $V_{\rm min}^{}$, ${\cal{O}}(T_{\rm TI}^{})$.  
%
As the Hubble scale during the thermal inflation, we have 
%We also define 
\aln{
H_{\rm TI}^{} =\frac{V_{\rm TI}^{1/2}}{\sqrt{3}M_{pl}^{}} ~.
\label{HTI}
}
%as a Hubble scale during the thermal inflation. 

At the QCD critical temperature $T_{\rm QCD}^{}\sim \Lambda_{\rm QCD}\sim 150$~MeV,
chiral condensation occurs and the Higgs potential acquires an additional linear term via the top Yukawa coupling \cite{Witten:1980ez}
%This QCD phase transition triggers the EW symmetry breaking via  the top yukawa term 
\aln{
 -\frac{y_t^{}}{\sqrt{2}}\langle \bar{t}t\rangle h\sim -\frac{y_t^{}}{\sqrt{2}}\Lambda_{\rm QCD}^3h~ .
}
Then the minimum of the Higgs field potential is shifted to have  a nonzero vev, 
\aln{
\langle h\rangle =v_{\rm QCD}^{}:= (y_t^{}\langle \bar{t}t\rangle/\sqrt{2})^{1/3}\sim \Lambda_{\rm QCD}^{}~.
\label{vQCD}
}
Simultaneously, axion field acquires a potential through QCD nonperturbative effects. 
We  now figure out why it is difficult to realize the axion-CMB scenario in the original $B$-$L$ model. 
As we discussed in the previous section, the lower bound of $R$ is ${\cal O}(0.01)$ due to the non-Gaussianity  constraints, 
and the Higgs vev in Eq.~(\ref{typical R}) must be ${\cal O}(100)$ Gev. Hence the above vev $\langle h \rangle \sim \Lambda_{\rm QCD}^{}$
is too small to be consistent with the non-Gaussianit constraint. 
In the next subsection, we will extend the model by introducing additional $O(N)$ scalar in order to evade this difficulty. 

Even after the Higgs acquires the QCD scale vev at $T=T_{\rm QCD}^{}$, 
the thermal inflation continues  because $\phi$ field remains  at $\phi=0$, 
if $m_{Z^\prime}^{}\geq \sqrt{2}m_H^{}$.
It can be seen by looking at the quadratic term of the finite temperature potential of $\phi$ in presence of $\langle h \rangle=v_{\rm QCD}^{}$;
\aln{
\frac{T^2M_{Z^\prime}^{}(\phi)^2}{8}-\frac{\lambda_{\phi H}^{}}{4}v_{\text{QCD}}^2\phi^2
&=\frac{m_{Z^\prime}^2}{8v_\phi^2}\left(T^2-2\left(\frac{m_H^{}}{m_{Z^\prime}}\right)^2 v_{\rm QCD}^2\right)\phi^2, 
\label{quadratic term in phi} 
}
where $m_H^{}=125$~GeV is the Higgs mass. 
If $m_{Z^\prime}^{}\geq \sqrt{2}m_H^{}$, the coefficient is positive at $T=T_{\rm QCD}^{}$, and
the scalar field is trapped at the false vacuum until the coefficient of quadratic term of $\phi$ becomes negative. 
The temperature at which the coefficient becomes negative and the thermal inflation 
ends\footnote{ 
As long as
\aln{
g_{B\hyphen L} > \frac{\sqrt{3}}{16 \pi}\frac{m_{Z^\prime}^3}{\Lambda_{\rm QCD} m_H^{} M_{pl}} \sim \paren{\frac{m_{Z^\prime}}{10 {\rm PeV}}}^3 ~, \label{SRV}
}
the slow-roll condition at $\langle h\rangle =v_{\rm QCD}$ is violated, so that $\phi$ starts rolling down towards the true minimum as soon as the temperature gets to $T_{\rm end}$.  
}
 is given by 
\aln{
T_{\rm end}^{}=\frac{m_H^{}}{\sqrt{2}m_{Z^\prime}^{}} \  v_{\rm QCD}^{}~. 
\label{Tend}
}
The e-folding number after $T=T_{\rm QCD}^{}$ 
%(\ref{efolding between QCD and end}) 
is now given by
\aln{
\Delta N_{\rm QCD}^{}
=\ln \left(\frac{T_{\rm QCD}^{}}{T_{\rm end}^{}}\right)=\ln\left( \frac{\sqrt{2}m_{Z^\prime}^{}}{m_H^{}}\right)~.
\label{e-folding L}
}
%___________________

%%%%%%%%%%%%%%%%%%%%%%%%%%%%%%%%%%%%%%%%%%%%%%%%%%%
%_________________________________
\subsection{$O(N)$ scalar extension of the CC $B$-$L$ model}
In the previous section, we have seen the difficulty of realizing the CMB anisotropy by QCD axion in the original  CC $B$-$L$ model due to the smallness of the Higgs vev at the time of QCD phase transition.  
In this section, we discuss a possibility to overcome this situation by utilizing the idea of the symmetry non-restoration (SNR)~\cite{Weinberg:1974hy}.  

We extend the model by adding a gauge singlet ${\cal O}(N)$ scalar field $S$;
\aln{
{\cal L}={\cal L}_{B-L}^{}+\frac{1}{2}(\partial S)^2 - V(S, H, \Phi)~.
}
Scalar potential is given by
\aln{
V(S, H, \Phi) =
\frac{m_S^2}{2}S^2 + \frac{\lambda_S^{}}{4!}(S^2)^2 - \frac{\lambda_{SH}^{}}{2}S^2(H^\dagger H)
+ \frac{\lambda_{S\phi}^{}}{2}S^2(\Phi^\dagger \Phi)~,
\label{model} 
}
where ${\cal L}_{B-L}^{}$ is the Lagrangian of the $B$-$L$ model with a QCD axion and $S^2:=\sum_{i=1}^N S_i^2$.  
In the following, we call a total set of SM, $B$-$L$ and QCD axion field as the {\it SM sector} to distinguish $S$ fields as
the {\it $S$ sector}.

Usually, the thermal mass of the SM Higgs, denoted by $\Pi_h$, is positive because of  the large positive contribution from the top-loop $\sim y_t^{2}T^2$. 
However, as discussed in Refs.~\cite{Weinberg:1974hy,Meade:2018saz,Baldes:2018nel,Glioti:2018roy,DAgnolo:2019cio,DiLuzio:2019wsw,Matsedonskyi:2020mlz,Chaudhuri:2020xxb}, it is also possible to obtain a negative effective mass  through a negative Higgs portal coupling $-\lambda_{SH}^{}$. 
Qualitatively, such a contribution is given by $\Delta \Pi_h^{}\sim -N\lambda_{SH}^{}\tilde{T}^2$, 
where $\tilde{T}$ is the temperature of the $S$ field. 
In the following, we use $\tilde{T}$ to denote the temperature of the $S$ sector while 
$T$ without a tilde denotes the temperature of the SM sector. 
It indicates that $\Pi_h^{}$ can be negative when the combination 
\aln{
\tilde{\lambda}_{SH}^{}:=N\lambda_{SH}^{}
}
is  sufficiently large. 
Apparently, the behavior of the Higgs thermal mass  can significantly change the thermal history 
of the universe and have various cosmological  implications.  
See Refs.~\cite{Meade:2018saz,Baldes:2018nel,Glioti:2018roy,DAgnolo:2019cio,DiLuzio:2019wsw,Matsedonskyi:2020mlz,Chaudhuri:2020xxb} %and references therein 
for recent studies. 
It is usually assumed that $O(N)$ scalars $S$ are in the thermal equilibrium at the same temperature as the SM particles, $\tilde{T}=T$.    
For our present purpose, however, 
it is not desirable since all the energy scale is then given by the QCD scale  at $\tilde{T}=T \sim T_{\rm QCD}$ and
 the resultant Higgs vev would be also given by the same  scale.  %$\sim T_{\rm QCD}^{}$ 
Hence  it is difficult to obtain $\langle h \rangle \sim 100$ GeV at   $T \sim T_{\rm QCD}^{}$.

In order to achieve a large Higgs vev,  we will consider a different thermal history in which
$S$ evolves almost separately from the SM sector in the early universe.    
If the temperature (or density) of $S$ is much higher than that of  the SM sector, $\tilde{T} \gg T$, 
we can have a large negative contribution to the effective Higgs mass, and 
consequently $\langle h \rangle \sim 100$ GeV even at the QCD phase transition.  

Based on the idea, we consider the following cosmological history of  Eq.~(\ref{model}):   
%%%%%%%%%%%%%%%%%%%%%%%%%%%%
%_____________________________________________
%%%%%%%%%%%%%%%%%%%%%%%%%%%
\begin{figure}[t!]
\begin{center}
\includegraphics[width=10cm]{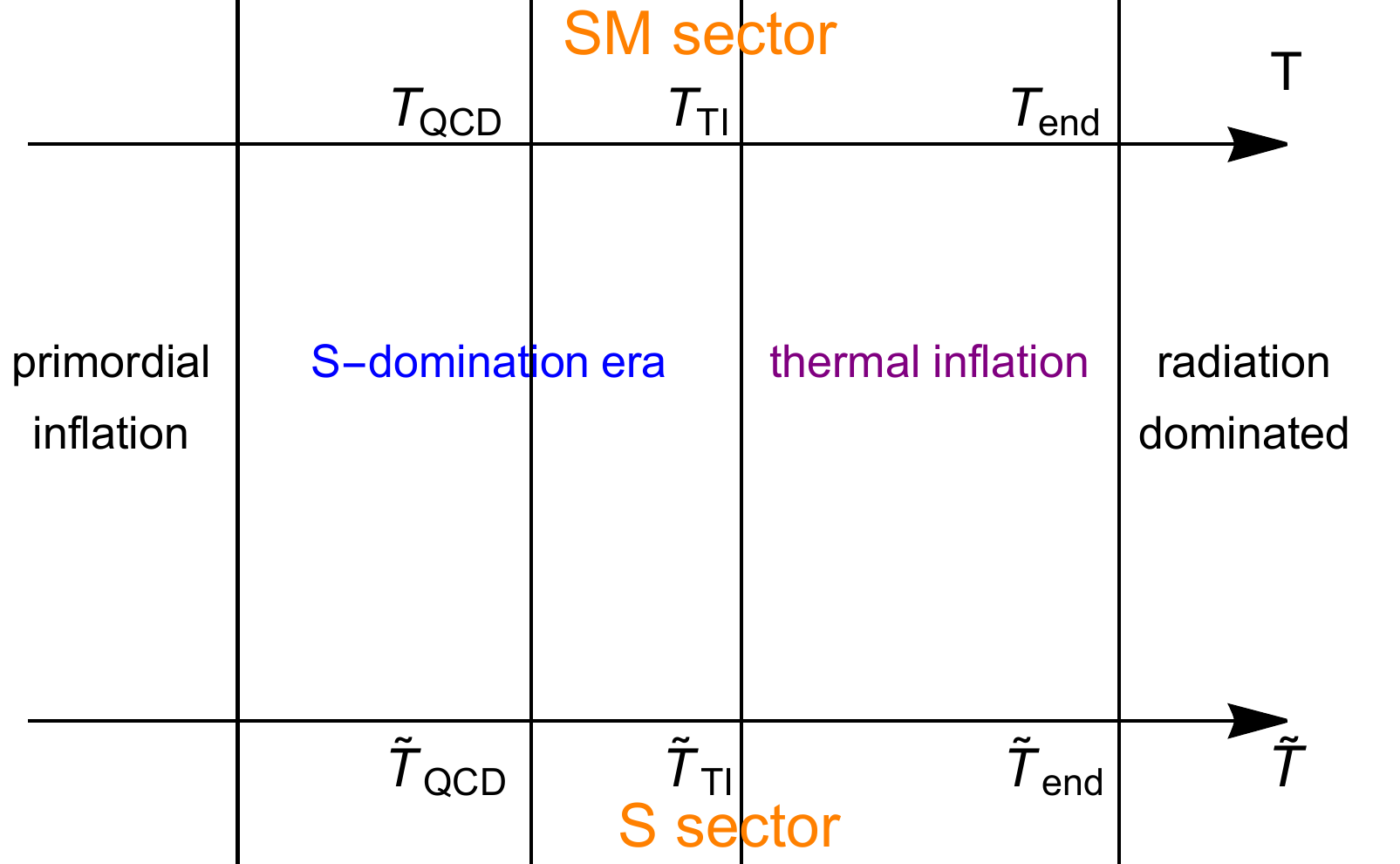}
\caption{
Thermal history of the early universe of the Axion-CMB scenario in the $O(N)$ extended model.  
The $O(N)$ sector evolves with different temperature $\tilde{T}$ 
from the SM sector with temperature $T$. 
}
\label{fig:history}
\end{center}
\end{figure}
%%%%%%%%%%%%%%%%%%%%%%%%%%%
\begin{enumerate}
\item $O(N)$ scalars $S$ are supposed to be dominantly produced in the primordial reheating and dominates the energy density of the universe until thermal inflation starts.~   
\\
We represent the energy densities and temperatures of ($S$, SM) sectors 
 as ($\rho_S^{}$,  $\rho_{\rm SM}^{}$) and ($\tilde{T}$, $T^{}$) respectively. 
\item Production  processes $SS\rightarrow hh$ and $\phi \phi$ are assumed 
to be very tiny so that the temperature of two sectors evolve differently.
% until  they become identical $T^{}=T_{\rm QCD}^{}$. 
%
The condition is  guaranteed as long as the scalar mixings, $\lambda_{SH}^{},\ \lambda_{S\phi}^{}$,  are sufficiently small.   
\item 
At  $T=T_{\rm QCD}^{}\sim 150 $ MeV, the axion acquires potential and the primordial fluctuation of the axion field is converted to the fluctuations of the SM radiation. 
The temperature of the $S$ sector is denoted by $\tilde{T}_{\rm QCD}^{}$ at this moment.
{After that, the production of the SM radiation from the $O(N)$ scalar should be ineffective so that the fluctuations are not diluted.}
\item Thermal inflation starts 
at $\tilde{T}=\tilde{T}_{\rm TI}^{}$ (and $T=T_{\rm TI}^{})$ when the dominant radiation energy $\rho_S^{}$ in the early universe becomes comparable to $V_{\rm TI}^{}$ of  Eq.~(\ref{vacuum energy}), and dilutes the  initial abundance of $S$. 
\item Thermal inflation ends when the trapped field $\phi$  starts rolling down at the temperature
 $\tilde{T}=\tilde{T}_{\rm end}^{}~(T=T_{\rm end}^{})$. 
%given in Eq.~(\ref{Tend}). 

\item Vacuum energy of the thermal inflation reheats up the SM sector.  
During and after the secondary reheating, the production of $S$ is required to be suppressed. 
%Due to our assumption $m_S^{}\gtrsim m_{Z^\prime}^{}$ and smallness of the scalar mixings $\lambda_{S \phi}$,  only the SM and $B$-$L$ particles are produced during the reheating process.  
The universe then follows the usual thermal history.  
\end{enumerate} 
%%%%%%%%%%%%%%%%%%%
Fig.~\ref{fig:history} is a schematic picture of the thermal history of the $O(N)$ extended model. 
The thermal history of our model is highly dependent on the mass scale of $S$ 
and we will study two extremal cases, $m_S \gg m_{Z^\prime}$ and $m_S=0$.
As we see, a viable model needs to satisfy the following two requirements:
\begin{itemize}
\item a requirement of  large Higgs vev  $\langle h\rangle \sim 246$ GeV at $T^{}=T_{\rm QCD}^{}$ gives a {constraint},  Eq.~(\ref{N in massive case}) for a massive case or Eq.~(\ref{N in massless case}) for a massless case, among various parameters of the model. 
\item 
a requirement of sufficient dilutions of $S$ and axions  during thermal inflation gives additional constraints  on the model. 
%
%Especially, $m_{Z^\prime}^{}$ is constrained because the end of thermal inflation is determined by $m_{Z^\prime}^{}$. % as shown in Eq.~(\ref{Tend}).  
%
\end{itemize}
As we see in the following sections, we will find that only the massless case can be consistent with all the theoretical and observational constraints for $N\gtrsim 10^{19}$ and $m_{Z'}^{}={\cal O}(10~{\rm TeV})$.

%%%%%%%%%%%%%%%%%%%%%%%%%%%%%%%%
%___________________________________________
\subsection{Massive $O(N)$ sector with $m_S \geq m_{Z^\prime}$}
We first study details of the thermal history in the massive $S$ case. 
Although we see that the massive case does not have an allowed parameter region for the axion-CMB scenario,
we will show the detailed calculations for comparison to the massless case.
We assume $m_S^{}\geq m_{Z'}^{}$ to forbid reproduction of $S$ after the $B$-$L$ thermal inflation.  
As depicted in Fig.~\ref{fig:history}, 
the cosmological history can be divided into three different eras: matter dominated era by $O(N)$ scalar, thermal inflation era, and then the ordinary SM radiation dominated era follows the reheating after the thermal inflation. 
\

\vspace{5mm}
%%%%%%%%%%%%%%%%%%%%%%%%%%%%%%%%%%%%%%
\noindent$\bullet$ {\bf Evolution of $\langle S^2 \rangle$ and $\langle h \rangle$
%Evolution of $O(N)$ scalar 
}\\
We assume that the energy density that has driven the primordial inflation is converted only to  the $O(N)$ scalar particle $S$, 
not to the SM sector particles. 
The initial distribution function of  $S$  is set when $S$ becomes decoupled from 
other unspecified fields such as inflatons, and given by
\aln{
f_{S,{\rm dec}}(k) = \paren{ e^{  (E_k - \mu_{\rm dec})/ \tilde{T}_{\rm dec}} -1}^{-1} \simeq e^{ (\mu_{\rm dec} - E_k )/ \tilde{T}_{\rm dec}} ~,
}
where $E_k = \sqrt{m_S^2 + \paren{k/a_{\rm dec}}^2}$ and we have assumed $ m_S- \mu_{\rm dec} \gg \tilde{T}_{\rm dec}$ for
the Boltzmann approximation. 
After the decoupling, the number density is given by
\aln{
n_S = \frac{N}{a^3}  \int\frac{d^3 k}{(2\pi)^3 } f_{S,{\rm dec}}(k) \simeq N 
 \alpha  \left(\frac{m_S \tilde{T}}{2\pi}\right)^{3/2}  ~, \label{kinetic density}
}
where we have defined $ \alpha := e^{ (\mu_{\rm dec} - m_S )/ \tilde{T}_{\rm dec}} $
and the effective temperature decreases as
 $\tilde{T} := \tilde{T}_{\rm dec}  (a_{\rm dec}/  a)^2$.
The number density $n_S$ and the energy density  $\rho_S \simeq m_S  n_S$
are diluted as $n_S, \rho_S  \propto a^{-3}$.

Thermal inflation starts at $t=t_{\rm TI}$ when the energy density of $S$, which has dominated until then,  becomes identical to the vacuum energy $ V_{\rm TI} $ of the $B$-$L$ field,
\aln{\rho_S \simeq m_S n_S = V_{\rm TI} ~. } 
Eq.~(\ref{kinetic density}) tells us that the effective temperature of the $S$ sector at this moment is 
\aln{\tilde{T}_{\rm TI} =  \frac{2\pi}{m_S^{5/3}( \alpha N)^{2/3}}V_{\rm TI}^{2/3}\sim 0.013~{\rm GeV}\left(\frac{m_{Z^\prime}^{}}{3\times 10^9~{\rm GeV}}\right)^{}\left(\frac{4\times 10^{15}}{\alpha N}\right)^{2/3}
%0.11\times \frac{m_{Z^\prime}^{}}{(\alpha N)^{2/3}}
\left(\frac{m_{Z^\prime}^{}}{m_S^{}}\right)^{5/3}
~. 
\label{temperature in massive case}
}
We denote the SM energy density at $t=t_{\rm TI}^{}$ as 
\aln{\rho_{\rm SM, TI}^{}:=\frac{\pi^2g_{\rm SM}^{}}{30}T_{\rm TI}^4~, }
where $T_{\rm TI}^{}$ is calculated below.  
In the following, we set the scale factor as $a=1$ at $t=t_{\rm TI}^{}$. 
Then the Hubble parameter evolves as 
\aln{
H=H_{\rm TI}^{}\times \begin{cases} a^{-3/2} & \text{for }a\leq 1,   \text{\ massive $S$ dominated}
\\
1 & \text{for }a>1,   \text{\ until the end of thermal inflation }
\end{cases}~,
}
where $H_{\rm TI}^{}$ is  defined by Eq.~(\ref{HTI}). 
The scale factor at the decoupling is given by
\aln{
a_{\rm dec} =\left(\tilde{T}_{\rm TI}^{}/\tilde{T}_{\rm dec}^{}\right)^{1/2}~,
}
which becomes very small if  decoupling occurs much earlier than the start of the thermal inflation.
Since $S$ is massive and behaves non-relativistically, 
the energy density $\rho_S$ is written in terms of the averaged amplitude of fluctuations as
%at finite density
\aln{
 \rho_S  \simeq m_S^2\langle S^2\rangle~.
 }
  Thus we have
\aln{
\langle S^2\rangle =\frac{V_{\rm TI}^{}}{m_S^{2}}a^{-3}=\frac{3m_{Z^\prime}^2}{128\pi^2 }\left(\frac{m_{Z^\prime}^{}}{m_S^{}}\right)^2  a^{-3}, 
\label{mean square value of S}
}
where  Eq.~(\ref{vacuum energy}) is used in the last equality. 

Through the portal coupling $\lambda_{SH}^{}$, the Higgs also acquires time-dependent vev in the early universe  as 
\aln{
\langle h\rangle =\sqrt{\frac{\lambda_{SH}^{}}{2\lambda_H^{}}\langle S^2\rangle }
\sim \frac{3^{1/2}m_{Z^\prime}^{}}{16\pi} \sqrt{\frac{\lambda_{SH}^{}}{\lambda_H^{}}} \left(\frac{m_{Z^\prime}^{}}{m_S^{}}\right) a^{-3/2}~.
%\ll m_{Z^\prime}^{}. 
\label{Higgs vev and S}
}
%

\begin{figure}[t!]
\begin{center}
\includegraphics[width=7cm]{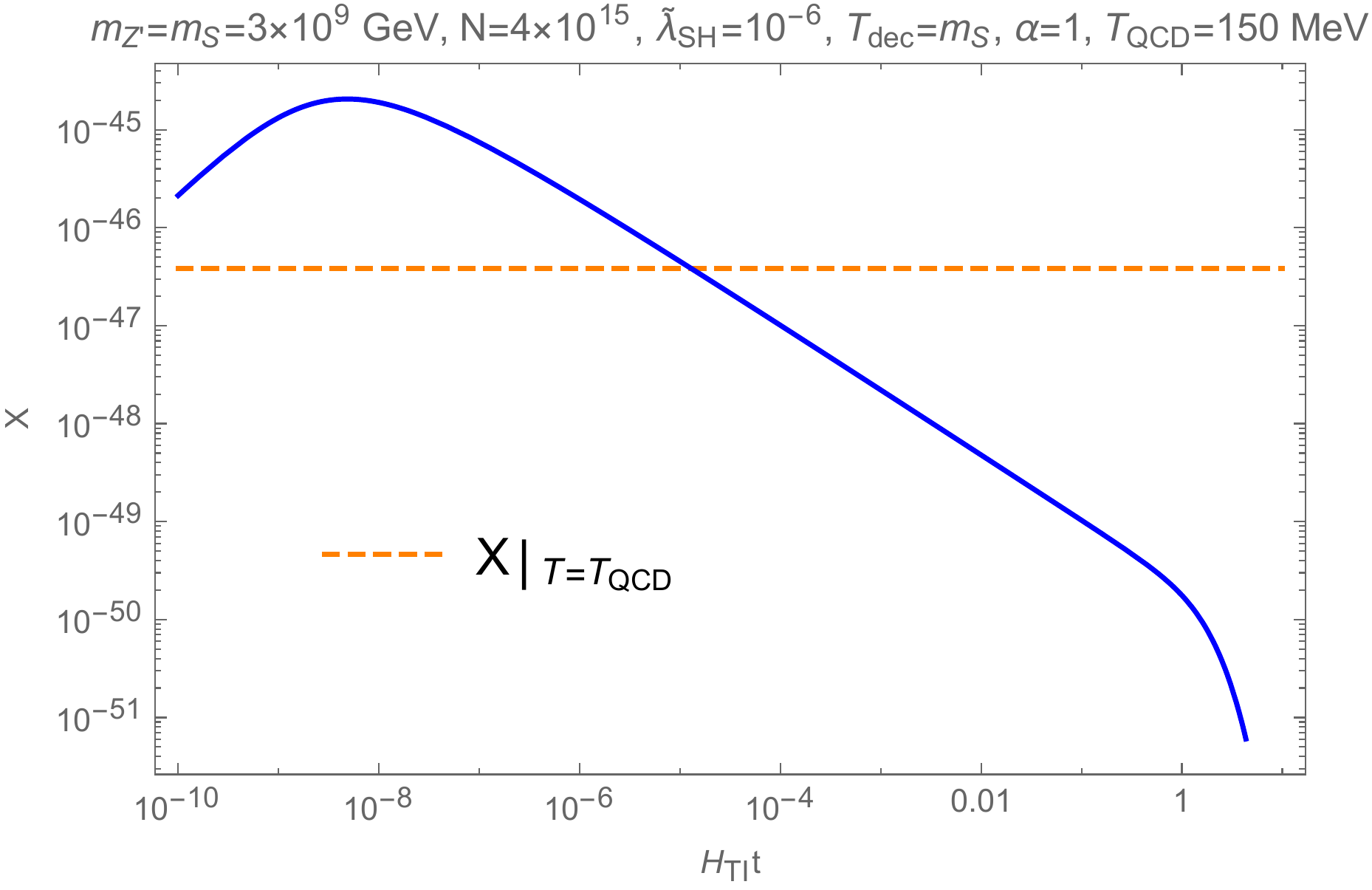}
\includegraphics[width=7cm]{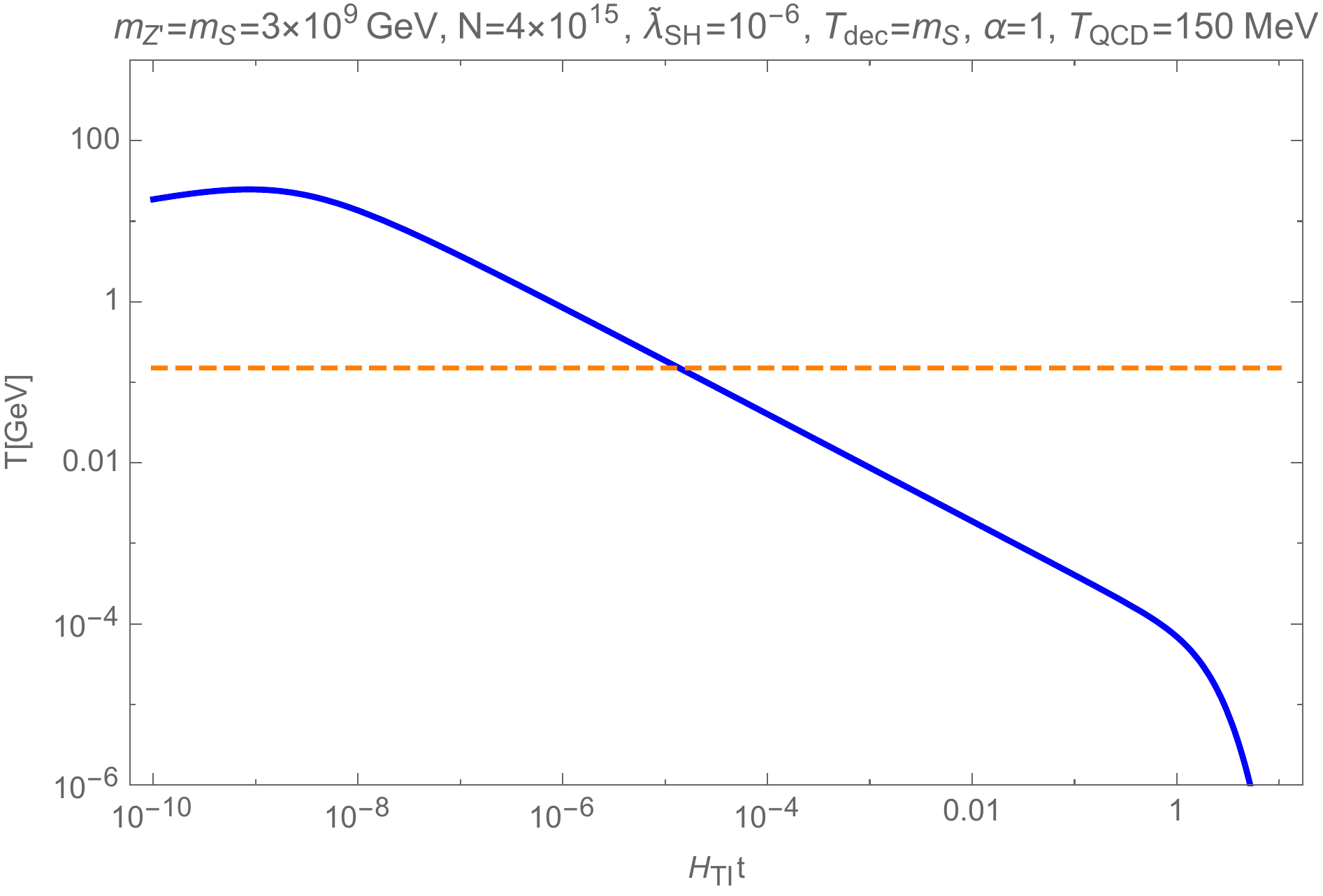}
\\
\includegraphics[width=7cm]{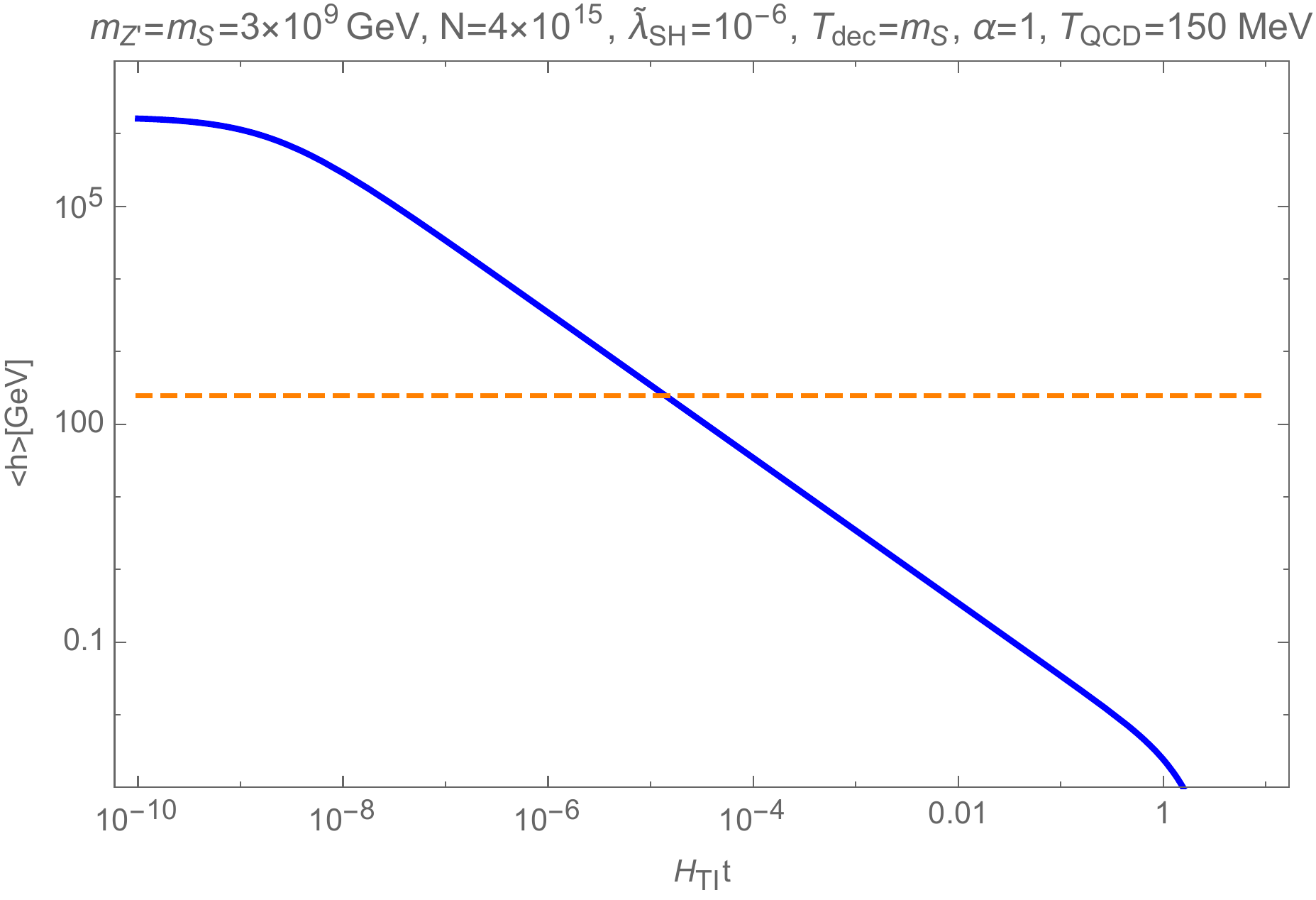}
\caption{
Time evolutions of the ratio $X=\rho_{\rm SM}/\rho_S$ (top-left), temperature of the SM sector $T$ (top-right) and the Higgs vev (down).  
The orange line corresponds to the  value of each quantity at $T=T_{\rm QCD}$. 
}
\label{figX}
\end{center}
\end{figure}

%%%%%%%%%%%%%%%%%%%
\vspace{5mm}
\noindent$\bullet$ {\bf Evolution of $\tilde{T}$} \\
\noindent
It is useful to rewrite $\rho_S^{}$ as a function of $\langle h\rangle$ by eliminating $\langle S^2\rangle$ from the above equations,
\aln{
\rho_S^{}=\frac{2\lambda_H^{}m_S^{2}\langle h\rangle ^2}{\lambda_{SH}^{}}~. 
\label{ns and v}
}
By equating this to $m_S^{}n_S^{}$ with Eq.~(\ref{kinetic density}), the evolution of the temperature $\tilde{T}$ can be  written in terms of $\langle h \rangle$ as
\aln{
\tilde{T} &= 2\pi\left(\frac{2\lambda_H^{}}{\alpha \tilde{\lambda}_{SH}^{}}\right)^{2/3}\frac{\langle h\rangle^{4/3}}{m_S^{1/3}}
\nonumber \\ 
&=  2.3
\times 10^{4}~{\rm GeV}\left(\frac{\lambda_H^{}}{0.1}\right)^{2/3}\left(\frac{10^{-6}}{\alpha \tilde{\lambda}_{SH}^{}}\right)^{2/3}\left(\frac{3\times 10^{9}~{\rm GeV}}{m_S^{}}\right)^{1/3}\left(\frac{\langle h\rangle }{246~{\rm GeV}}\right)^{4/3}~.
\label{tildeT and h}
}
See the schematic picture in Fig.~\ref{fig:history} and the
lower panel of Fig.~\ref{figX}. 
The induced Higgs mass via $\langle S^2 \rangle$ is given by
$m_{H,{\rm eff}} =  \sqrt{\lambda_{SH}  \langle S^2 \rangle }$.  Then the ratio of $m_S$ to this 
becomes 
\aln{
\left(\frac{m_S^{}}{m_{H,{\rm eff}}}\right)^2=\frac{1}{\tilde{\lambda}_{SH} \alpha}\left(\frac{2\pi m_S^{}}{T}\right)^{3/2}\gg 1~, 
\label{mass ratio}
}
and thus $hh \rightarrow SS$ is kinematically suppressed so that
we can safely neglect  the process. 

\

%%%%%%%%%%%%%%%%%%%%%%%%%%%%%%%%%%%%%
\noindent$\bullet$ {\bf Production of SM particles and evolution of $X=\rho_{\rm SM}/ \rho_S$
}\\
The  SM radiation $\rho_{\rm SM}$
is generated  by the $SS\to HH$ process with the  weak $\lambda_{SH}$ coupling. 
Let us now study its time evolution. 
Our assumption is that the process is extremely low and the ratio
\aln{
X:=\rho_{\rm SM}^{}/\rho_S^{} \label{def of X}
}
is much less that unity until the thermal inflation. 
Under this assumption, we can safely neglect the back-reaction to $\rho_S^{}$ and the behavior $\rho_S^{} \propto a^{-3}$ discussed above remains intact. 
Then the Boltzmann equation of $\rho_{\rm SM}$ is given by
\aln{
\dot{\rho}_{\rm SM}^{}+4 H\rho_{\rm SM} \simeq m_S  \frac{\lambda_{SH}^2}{16\pi m_S^2}n_S^{}n_i^{} 
= \frac{\lambda_{SH}^2}{16\pi m_S^3}\rho_S \rho_i , 
\label{Boltzmann equation of nSM}
}
where $n_i:=n_S/N$ and $\rho_i:=\rho_S/N$.    
Its derivation  is presented in \ref{app:Boltzmann}. 
From this, the ratio (\ref{def of X}) obeys the following equation;
\aln{
&\frac{dX}{dt} + H X = \frac{\lambda_{SH}^2}{16 \pi Nm_S^3}\rho_S 
 =\frac{\lambda_{SH}^2 V_{\rm TI}}{16\pi Nm_S^3}a^{-3} ~.
}
Inserting the explicit time-dependence of $\rho_S$ and the Hubble parameter $H=d \ln a/dt$, we have
\aln{
 & \frac{d(aX)}{da}=\frac{\lambda_{SH}^2 V_{\rm TI}}{16\pi N m_S^3 H_{}^{}}a^{-3}:=c\times \begin{cases}a^{-3/2} &{\rm for}\ a\leq 1
\\
a^{-3} &{\rm for}\ a>1
\end{cases},
\label{equation of X}
}
where
\aln{
c :=\frac{\sqrt{3}\tilde{\lambda}_{SH}^2M_{pl}^{}V_{\rm TI}^{1/2}}{16\pi N^3m_S^{3}}
~. 
}
It can be integrated as (noting $a_{\rm TI}=1$)
\aln{
aX-X_{\rm TI}^{}=c\times \begin{cases}2(1-a^{-1/2}) & {\rm for}\ a\leq 1
\\
\frac{1}{2}(1-a^{-2})&{\rm for}\ a>1 
\end{cases}~,
}
where $X_{\rm TI}^{}$ is the value of $X$ at $T=T_{\rm TI}^{}$.  
By denoting the initial value of $X$ right after the decoupling of $S$ as $X_{\rm dec}^{}$, we have
\aln{
X_{\rm TI}=a_{\rm dec} X_{\rm dec} - 2c(1-a_{\rm dec}^{-1/2})\simeq a_{\rm dec} X_{\rm dec} + 2c a_{\rm dec}^{-1/2} ~.
  \label{X_TI}
}
%%%%%%%%%%%%%%%%%%%%%%%%%%%
%%%%%%%%%%%%%%%%%%%%%%%%%%%
The top-left figure of Fig.~\ref{figX} shows the time evolution of the ratio with $X_{\rm dec}^{} = 0$.  
It first grows with time through the energy transfer from $S$ and SM sectors. 
At  $a \sim (3c/X_{\rm TI})^2\sim (3/2)^2a_{\rm dec}^{}\ll 1$, it starts decreasing as $X \sim a^{-1}$ since the $SS\to HH$ process is not fast enough to overwhelm the expansion and $\rho_{\rm SM} \sim a^{-4}$ is diluted more quickly than $\rho_S^{} \sim a^{-3}$ is.
This behavior continues in the thermal inflation period;
\aln{
X  \rightarrow  \frac{1}{a}\left(X_{\rm TI}+ \frac{c}{2} \right)
\sim  \frac{X_{\rm TI} }{ a}   \quad \text{for }a\gg 1~.
\label{X solution in massive case}
}
In the following, it is supposed %that $a_{\rm dec}$ is sufficiently small
that the first term $a_{\rm dec} X_{\rm dec}$ in Eq.~(\ref{X_TI}), which depends  on the unspecified dynamics before the decoupling, is negligible.
The initial conditions of later evolution are given by
 $\tilde{T}_{\rm dec}$ and $\mu_{\rm dec}$,  which  characterize the physics in the primordial reheating era.

%%%%%%%%%%%%%%%%%%%%%%%%%%%%%%%%%%%%%%%%%%%%%%%%%%%%
\vspace{5mm}
\noindent$\bullet$ {\bf Various quantities at $T=T_{\rm QCD}^{}$} \\
\noindent
The top-right figure of Fig.~\ref{figX} is the temperature evolution of the SM sector, $T$. 
It first grows by the energy transfer from $S$ sector, but then 
decreases due to the expansion of the universe and eventually reaches $T_{\rm QCD}$, which is depicted 
by the orange line. 
The scale factor at which  $T$ gets down to $T_{\rm QCD} \sim  150~{\rm MeV}$ is evaluated as
\aln{
a_{\rm QCD}^{} &=\paren{\frac{\rho_{\rm SM, TI} }{\rho_{\rm SM, QCD}}}^{1/4} = \paren{\frac{X_{\rm TI} V_{\rm TI} }{\rho_{\rm SM, QCD}}}^{1/4} \label{aQCD in massive, 1st line}
\\
&=7.6\times 10^{-4}\times \alpha^{\frac{1}{24}}\paren{\frac{\tilde{\lambda}_{SH}^{}}{10^{-6}}}^{\frac{1}{2}}
\paren{\frac{4\times 10^{15}}{N}}^{17/24}
\paren{\frac{100}{g_{\rm SM}^{}}}^{1/4}
\paren{\frac{150~{\rm MeV}}{T_{\rm QCD}^{}}}
\nonumber \\
& \quad \quad \times
\paren{\frac{m_{Z'}^{}}{3\times 10^9~{\rm GeV}}}^{\frac{3}{4}}
\paren{\frac{m_{Z'}^{}}{m_S^{}}}^{\frac{7}{12}} \paren{\frac{\tilde{T}_{\rm dec}}{m_S}}^{1/16}  
~, 
\label{aQCD in massive}
}
where we used $\rho_{\rm SM, QCD} = (g_{\rm SM} \pi^2 / 30) T_{\rm QCD}^4$. 
From Eq.~(\ref{aQCD in massive}),  
we can confirm  $\tilde{T}_{\rm QCD}^{}=\tilde{T}_{\rm TI}^{}a_{\rm QCD}^{-1}\gg \tilde{T}_{\rm TI}^{}$ again. 
As mentioned after Eq.~(\ref{tildeT and h}), 
 $T=T_{\rm QCD}^{}$ is realized well before the beginning of thermal inflation.   
Note also that the SM temperature $T_{\rm TI}^{}$ satisfies
\aln{T_{\rm TI}^{}=T_{\rm QCD}^{}a_{\rm QCD}^{}\ll T_{\rm QCD}^{}~. 
}

Recall that we need $\langle h \rangle={\cal O}(100\ {\rm GeV})$ at $T^{}=T_{\rm QCD}^{}$ for the axion-CMB scenario to be observationally viable  in the CC $B$-$L$ model.  
The time evolution of $\langle h \rangle$  is plotted in the down figure of Fig.~\ref{figX}. 
From Eqs.~(\ref{ns and v}) and (\ref{def of X}), the condition  for the Higgs vev is written as  
\aln{
X|_{T^{}=T_{\rm QCD}^{}} %=\frac{2\zeta(3)g_{\rm SM}^{}\lambda_{SH}^{}T_{\rm QCD}^3}{\pi^2 \lambda_H^{}m_S^{}\langle h\rangle^2|_{T^{}=T_{\rm QCD}^{}}}~. 
= \frac{ \lambda_{SH}^{} \rho_{\rm SM, QCD}^{}}{2 \lambda_H^{} m_S^2 \langle h\rangle^2|_{T^{}=T_{\rm QCD}^{}}}~. 
\label{required X}
}
On the other hand, from Eqs.~(\ref{X solution in massive case}) and (\ref{aQCD in massive, 1st line}), we have
\aln{
X|_{T^{}=T_{\rm QCD}^{}} = X_{\rm TI}^{3/4} \paren{\frac{\rho_{\rm SM, QCD}}{V_{\rm TI} } }^{1/4} ~. 
\label{solution of X}
}
By equating Eq.~(\ref{solution of X}) to Eq.~(\ref{required X}), $N$ is determined in terms of other parameters of the model; 
\aln{
N&\sim 4\times 10^{15}
\times \left(\frac{\tilde{\lambda}_{SH}^{}}{10^{-6}}\right)^{4/9}\paren{\frac{\lambda_H}{0.1}}^{8/9} \paren{\frac{100}{g_{\rm SM}}}^{2/3}    \paren{\frac{\langle h\rangle}{246 \ {\rm GeV}}}^{16/9} \paren{\frac{150 {\rm MeV}}{T_{\rm QCD}}}^{8/3} 
\nn
&\h{6cm}\times \paren{\frac{m_S^{}}{3\times 10^9~{\rm GeV}}}^{2/9}\paren{\frac{T_{\rm dec}^{}}{m_S^{}}}^{1/6}~. 
\label{N in massive case}
}
As we discussed in Eq.~(\ref{typical R}), the Higgs must have as large vev at $T=T_{\rm QCD}$ as the current value so that  it is consistent with the non-Gaussianity constraint
and the CMB amplitude can be explained in terms of the primordial axion fluctuations. 
Then, we see from the above relation, that $N$ must be quite large. 

Finally note that the assumption $a_{\rm dec} \ll 1$ is justified as far as
\aln{
\paren{\frac{T_{\rm dec}}{m_S}}^{3/2} \alpha \times N \gg (2 \pi)^{3/2} \frac{V_{\rm TI}}{m_S^4} ~.
}

\vspace{5mm}
%%%%%%%%%%%%%%%%%%%%%%%%%%%%%%%%
\noindent$\bullet$ {\bf Dilution factor by thermal inflation}\\
By definition, the number density of $S$ at $\tilde{T}=\tilde{T}_{\rm TI}^{}$ is $V_{\rm TI}^{}/m_S^{}$ and it is diluted rapidly during the thermal inflation. 
The Higgs vev $\langle h\rangle$ decreases and finally approaches the value $v_{\rm QCD}$ determined by the QCD chiral condensates in Eq.~(\ref{vQCD});
\aln{
\langle h\rangle^2=\begin{cases}\langle h\rangle^2|_{T^{}=T_{\rm QCD}}^{}\left(a_{\rm QCD}^{}/a\right)^3 & T^{}\sim T_{\rm QCD}^{}
\\
v_{\rm QCD}^2 & T^{}\ll T_{\rm QCD}^{}
\end{cases}~.  
%\frac{2\tilde{\lambda}_{SH}^{}}{N\lambda_H^{}m_S^{}}\times n_S^{{\rm QCD}}\left(\frac{a_{\rm QCD}^{}}{a}\right)^3,
\label{vev after TQCD}
}
where  $a_{\rm QCD}^{}$ is given by Eq.~(\ref{aQCD in massive}).     

To calculate the dilution factor, we need to know the temperature $T_{\rm end}^{}$ at which the thermal inflation ends. 
In presence of the $S$ sector, the quadratic term of the $B$-$L$ scalar $\phi$ of Eq.~(\ref{quadratic term in phi}) is modified to be
\aln{
&  \frac{T^2}{8} 
M_{Z^\prime}^2(\phi)-\frac{\lambda_{\phi H}^{}}{4}\langle h\rangle^2 \phi^2+
\frac{\tilde{\lambda}_{S\phi}^{}}{4N}\langle S^2\rangle \phi^2~. 
\label{rollingpotential}
}
For simplicity, we consider a situation where the last term is negligibly small by assuming the smallness of $\tilde{\lambda}_{S\phi}^{}$,
\aln{
\frac{\tilde{\lambda}_{S\phi}^{}}{4N}\langle S^2\rangle = 
\frac{\tilde{\lambda}_{S\phi}^{}\lambda_H^{}}{2\tilde{\lambda}_{SH}^{}}\langle h\rangle^2 
\ll \frac{\lambda_{\phi H}^{}}{4}\langle h\rangle^2~,
}
where we have used Eq.~(\ref{Higgs vev and S}).  
Then the end of the thermal inflation is determined by the first two terms of Eq.~(\ref{rollingpotential});
\aln{
&\frac{T^2}{2}g_{B-L}^2\phi^2-g_{B-L}^2\left(\frac{m_h^{}}{m_{Z^\prime}^{}}\right)^2\langle h\rangle^2 \phi^2
%+ \frac{\tilde{\lambda}_{S\phi}^{}\lambda_H^{}}{2\tilde{\lambda}_{SH}^{}}\langle h\rangle^2 \phi^2
= g_{B-L}^2\left[\frac{T^2}{2}-\left(\frac{m_H^{}}{m_{Z^\prime}^{}}\right)^2\langle h\rangle^2 \right]\phi^2~. 
%+\frac{\tilde{\lambda}_{S\phi}^{}\lambda_H^{}}{2\tilde{\lambda}_{SH}^{}}\langle h\rangle^2 \phi^2~,
\label{quadratic of phi}
}
As long as $\langle h\rangle^2$ decreases as $a^{-3}$ (see Eq.~(\ref{vev after TQCD})), the first term in Eq.~(\ref{quadratic of phi}) is always dominant because $T^{2}$ decreases as $a^{-2}$.  
Thus, these two terms can become comparable after $\langle h\rangle$ reaches the constant value $v_{\rm QCD}^{}$, which means that $T_{\rm end}^{}$ is similarly determined as in the case of the conventional $B$-$L$ model and given by Eq.~(\ref{Tend}).   
Then the dilution factor during the thermal inflation 
 from $T=T_{\rm TI}^{}$ until $T=T_{\rm end}^{}$ is obtained as
\aln{
e^{-3\Delta N}=\left(\frac{T_{\rm end}^{}}{T_{\rm TI}^{}}\right)^3=\frac{1}{2^{3/2}}\left(\frac{v_{\rm QCD}^{}}{T_{\rm QCD}^{}}\right)^3\left(\frac{m_H^{}}{m_{Z^\prime}^{}}\right)^3\times a_{\rm QCD}^{-3}~. 
\label{dilution in massive case}
}
If $m_{Z^\prime}$ is sufficiently heavy, the dilution factor during the thermal inflation becomes large.

\vspace{5mm}
%%%%%%%%%%%%%%%%%%%%%%%%%%%%
\noindent$\bullet$ {\bf Dilution of axions}\\
We now estimate $r_A^{}$ in the massive case. 
The axion field starts to oscillate when $m_A^{}(T)\geq 3H$ is satisfied. 
Here, $m_A^{}(T)$ is the temperature-dependent axion mass: 
\aln{
m_A^{}(T)=m_{A0}^{}\times \begin{cases}(T_{\rm QCD}^{}/T)^{4b}& \text{for $T\geq T_{\rm QCD}^{}$}
\\
1  & \text{for $T\leq T_{\rm QCD}^{}$}
\end{cases}~,
}
where $b\sim 1.02$. 
In the massive $S$ case, the solution is determined by $m_{A0}^{}=3H$;
\aln{
\tilde{T}_{\rm osc}^{}=\frac{2\pi(M_{pl}^{}m_{A0}^{})^{4/3}}{3^{2/3}(\alpha N)^{2/3}m_S^{5/3}}
=6.9\times 10^{-21}~{\rm GeV}\left(\frac{4\times 10^{15}}{\alpha N}\right)^{2/3}\left(\frac{3\times 10^{9}~{\rm GeV}}{m_S^{}}\right)^{5/3}\left(\frac{m_{A0}^{}}{6\times 10^{-6}~{\rm eV}}\right)^{4/3}~,  
\label{Tosc in massless}
}
which is actually tiny compared to $\tilde{T}_{\rm QCD}^{}$ and $\tilde{T}_{\rm TI}^{}$, which is consistent with the assumption $T_{\rm osc}^{}<T_{\rm QCD}^{}$. 
Thus, the axion field does not evolve until the beginning of thermal inflation and we can use the same result of $r_A^{}$  calculated in Ref.~\cite{Iso:2020pzv} with a different e-folding number; 
%\mage{(thermal inflationが始まる前も始まった後もslow-rollであるという意味では、thermal inflationの前の転がりも一応考慮すべきでしょうか。)}
\aln{
r_A^{} \sim \paren{\frac{2 \tan \paren{\bar{\theta}_{\rm ini}/2}}{0.3}}^2 \paren{\frac{f_A}{10^{12} {\rm GeV}} }^{1.16} \exp \paren{-2 \eta \times \Delta N} , 
\label{r_A with thermal inflation1} 
}
where 
\aln{
\eta :=\frac{1}{3} \paren{\frac{m_{A0}}{H_{\rm TI}}}^2=8.6\times 10^{-6}\times \left(\frac{10^{12}~{\rm GeV}}{f_A^{}}\right)^2\left(\frac{10~{\rm TeV}}{m_{Z'}^{}}\right)^4~
 \label{axion slow-roll parameter}
}
and $ \Delta N=\log \left(T_{\rm TI}^{}/T_{\rm end}^{}\right)$. 
The result (\ref{r_A with thermal inflation1}) shows that,   in the massive case, 
$Z^\prime$ gauge boson has to be  lighter than 10 TeV  for
 realizing the necessary condition of  $r_A^{}\lesssim 10^{-4}$.   
On the other hand,  as we see in the following, a condition for sufficient dilution of $S$ requires $m_{Z^\prime} \gg 10$~TeV. 
% heavy $m_{Z^\prime} \gtrsim 10^6$~TeV.
 Thus it is impossible to satisfy all the observational constraints in the massive $S$ case. 

\

%%%%%%%%%%%%%%%%%%%%%%%%%%%%%%%%%%%%%%%%%%%%%%%%%
\noindent$\bullet$ {\bf Dilution of initial abundance of $S$}\\
In addition to the dilution of the axion abundance, the thermal inflation also has to dilute the initial abundance of $S$. 
Otherwise, such an abundance may cause various cosmological problems. 
Especially $r_S^{}:=\rho_S^{}/\rho_{\rm DM}^{}$ must satisfy the same constraint as in Eq.~(\ref{r and R}),
\aln{
r_S^{} < 8.2 \times 10^{-3} R~. 
\label{rS and R}
}
Let us now evaluate the relic abundance  of $S$ at the present universe.  
After the thermal inflation ends at $a=a_{\rm end}$, 
the energy density of the $\phi$-oscillation dominates the universe until the universe is reheated to the temperature $T_{R}$. 
%completion of the reheating at which the SM sector's temperature is $T_{R}$.
The relic abundance of $S$ is  estimated as 
\aln{
\rho_{S}^{}|_{\rm today}^{}&\sim 
\left(\frac{a_{\rm end}^{}}{a_{\rm today}}\right)^3 e^{-3\Delta N}\times V_{\rm TI}^{}%m_S^{}n_S^{\rm QCD}%n_S^{}|_{T^{}=T_{\rm QCD}^{}}^{}
\nn
&\sim 
\frac{g_R \pi^2}{30}\frac{T_{R}^4}{V_{\rm TI}} \left(\frac{T_{\rm today}^{}}{T_{R}}\right)^3e^{-3\Delta N}\times V_{\rm TI}^{}~.
%m_S^{}n_S^{\rm QCD}~,
 %|_{T^{}=T_{\rm QCD}^{}}^{}. 
\label{rhos now}
}
where  $T_{\rm today}^{}=2.73$ K and $g_R$ is the effective degrees of freedom at the reheating.  
By substituting Eq.~(\ref{dilution in massive case}) into Eq.~(\ref{rhos now}), we obtain 
\aln{
r_S^{} &:= \rho_{S}^{}/\rho_{\rm DM}^{}|_{\rm today}^{} =(\rho_{S}^{}/\rho_\gamma^{})(\rho_\gamma^{}/\rho_{\rm DM}^{})|_{\rm today}^{}
\nn
&=0.011\times \paren{\frac{10^{-6}}{\tilde{\lambda}_{SH}^{}}}^{3/2}
\paren{\frac{N}{4\times 10^{15}}}^{17/8}
\paren{\frac{v_{\rm QCD}^{}}{T_{\rm QCD}^{}}}^3\paren{\frac{T_{\rm QCD}^{}}{150~{\rm MeV}}}^3
\nn
& \times  \paren{\frac{10^6~{\rm TeV}}{m_{Z'}^{}}}^{21/4}
 \paren{\frac{m_S^{}}{m_{Z'}^{}}}^{7/4}
\paren{\frac{T_R^{}}{10~{\rm MeV}}} \paren{\frac{m_S^{}}{T_{\rm dec}^{}}}^{3/16}~,
 \label{present density}
}
where we  used $(\rho_\gamma^{}/\rho_{\rm DM}^{})|_{\rm today}^{}=2.9\times 10^{-4}$. 
Even with the reheating temperature around its lower limit $\sim 10$~MeV for the successful Big Bang nucleosynthesis, 
 large $m_{Z'}$ is required.
Therefore, the two necessary conditions of $r_A, r_S < 10^{-3}$ cannot be simultaneously satisfied.

\
 
 %%%%%%%%%%%%%%%%%%%%%%%%%%%
 %______________________massless case_____________________________
\subsection{Massless $O(N)$ sector}
In the previous subsection, we saw a difficulty to realize the axion-CMB scenario for   $m_S \ge m_{Z^\prime}$.     
Here, let us consider another extremal case of $m_S=0$. 
The following discussion is almost  parallel to the massive case 
except that (i) the SM production processes can be kinematically suppressed if $\tilde{\lambda}_{SH} \gtrsim 1$ and
 (ii) the SM radiation produced directly from the primordial reheating needs to be taken into account.   
%
%In Fig.~\ref{history_massless}, we show a schematic picture which describes the thermal history of massless case. 

\

%%%%%%%%%%%%%%%%%%%%%%%%%%%%%%%%%%%%%% 
\noindent$\bullet$ {\bf Evolution of $\langle h \rangle$}\\
The number and energy densities of $S$ after the primordial reheating are now given by 
\aln{
n_S^{} =\frac{N\zeta(3)}{\pi^2}\tilde{T}^3~,\quad 
\rho_S^{} =\frac{N\pi^2}{30}\tilde{T}^4~,  \label{massless density}
}
where $\tilde{T}= \tilde{T}_{\rm reh}^{}  (a_{\rm reh}/a)$. 
%
%(\blue{including $\mu$ ?})

Thermal inflation starts at $t=t_{\rm TI}$ when $\rho_S^{}$ becomes identical to the vacuum energy of the $B$-$L$ field, $\rho_S = V_{\rm TI}~$. 
The corresponding temperature of the $S$ sector is 
\aln{
\tilde{T}_{\rm TI} =  \left(\frac{30}{\pi^2 N}\right)^{1/4}V_{\rm TI}^{1/4}\sim 0.05~{\rm GeV}\times \left(\frac{m_{Z'}^{}}{10~{\rm TeV}}\right)\left(\frac{%9\times 10^{15}
10^{19}}{N}\right)^{1/4}~. 
\label{temperature in massless case}
}
Since the scale factor is set as $a=1$ at $t=t_{\rm TI}^{}$, 
 the Hubble parameter behaves as 
\aln{
H=H_{\rm TI}^{}\times \begin{cases} a^{-2} & \text{for }a\leq 1, \text{\ massless $S$ dominated }
\\
1 & \text{for }a>1, \text{\ until the end of thermal inflation }
\end{cases}~,
}
where $H_{\rm TI}^{}$ is  defined by Eq.~(\ref{HTI}). 
In the massless case, the Higgs vev is determined by the thermal mass $-\tilde{\lambda}_{\rm SH}^{}\tilde{T}^2/24$ as in the conventional studies \cite{Weinberg:1974hy,Meade:2018saz,Baldes:2018nel,Glioti:2018roy} and given by
\aln{
\langle h\rangle^2=\tilde{\lambda}_{SH}^{}\tilde{T}^2/(24\lambda_H^{})~.  
\label{vev in massless case}
}
We require that $\langle h \rangle \sim 246$  GeV at $T=T_{\rm QCD}^{}$.
Note also that, because of the Higgs vev, one of the components of $O(N)$ scalar acquires 
 mass $\sqrt{\lambda_{SH}^{}} \langle h \rangle \sim \tilde{\lambda}_{SH} \tilde{T} /\sqrt{N}$. 
 However, it is much smaller than $\tilde{T}$ so that the massive component can be regarded as practically massless.

\

%%%%%%%%%%%%%%%%%%%%%%%%%%%%%%%%%%%%%
\noindent$\bullet$ {\bf Production of SM particles and evolution of $X=\rho_{\rm SM}/ \rho_S^{}$
}\\
If $\lambda_{SH}^{} < 1$, 
$SS\to hh$ is the dominant process as in the massive case. 
On the other hand, for $\lambda_{SH}^{} \gtrsim 1$,  
the thermal mass of the Higgs $\sim \sqrt{\tilde{\lambda}_{SH}}\tilde{T}$ becomes
 larger than the radiation temperature $\widetilde{T}$ of the $O(N)$ scalar sector, 
 and the SM production rate gets Boltzmann-suppressed.
%\red{so are the other SM particle masses due to the Higgs vev (\ref{vev in massless case}). これは本当？}
Then, by introducing the suppression factor $\epsilon$, 
 the Boltzmann equation of $\rho_{\rm SM}^{}$ is given by
%the Boltzmann equation of $\rho_{\rm SM}^{}$ becomes
\aln{
\dot{\rho}_{\rm SM}^{}+4H\rho_{\rm SM}^{}&=\frac{\epsilon \lambda_{SH}^2}{128\pi\tilde{T}}n_S^{}n_i^{}~
\nn
&=\frac{{\epsilon} \tilde{\lambda}_{SH}^2}{128\pi N^{9/4}}\left(\frac{30}{\pi^2}\right)^{5/4}\left(\frac{\zeta(3)}{\pi^2}\right)^{2}\rho_S^{5/4}~,
}
where we have  used
\aln{
n_S^{}=\frac{30\zeta(3)}{\pi^4}\frac{\rho_S^{}}{\tilde{T}}~,\quad \tilde{T}=\left(\frac{30\rho_S^{}}{\pi^2 N}\right)^{1/4} ~. 
}
The suppression factor ${\epsilon} \leq 1$ is normalized as 
${\epsilon} = 1$ when the process $SS \to hh$ is  unsuppressed, and 
it dose not have any temperature dependence since there is no explicit mass scale.
Note that  SM particles other than Higgs are also produced; $SS \to \text{(SM particles)}$.
A process like $SS\to b\bar{b}$ may not be thermally suppressed unlike $SS\to hh$,
but  it is negligibly small  because of the tiny Yukawa coupling. For example 
 we have {$\epsilon \sim y_b^2 \tilde{\lambda}_{SH}^{-1} \sim 10^{-5}$ } for $\tilde{\lambda}_{SH} \sim 10$ %(to be checked!)
 and we can safely neglect them.

\begin{figure}[t!]
\begin{center}
\includegraphics[width=7.5cm]{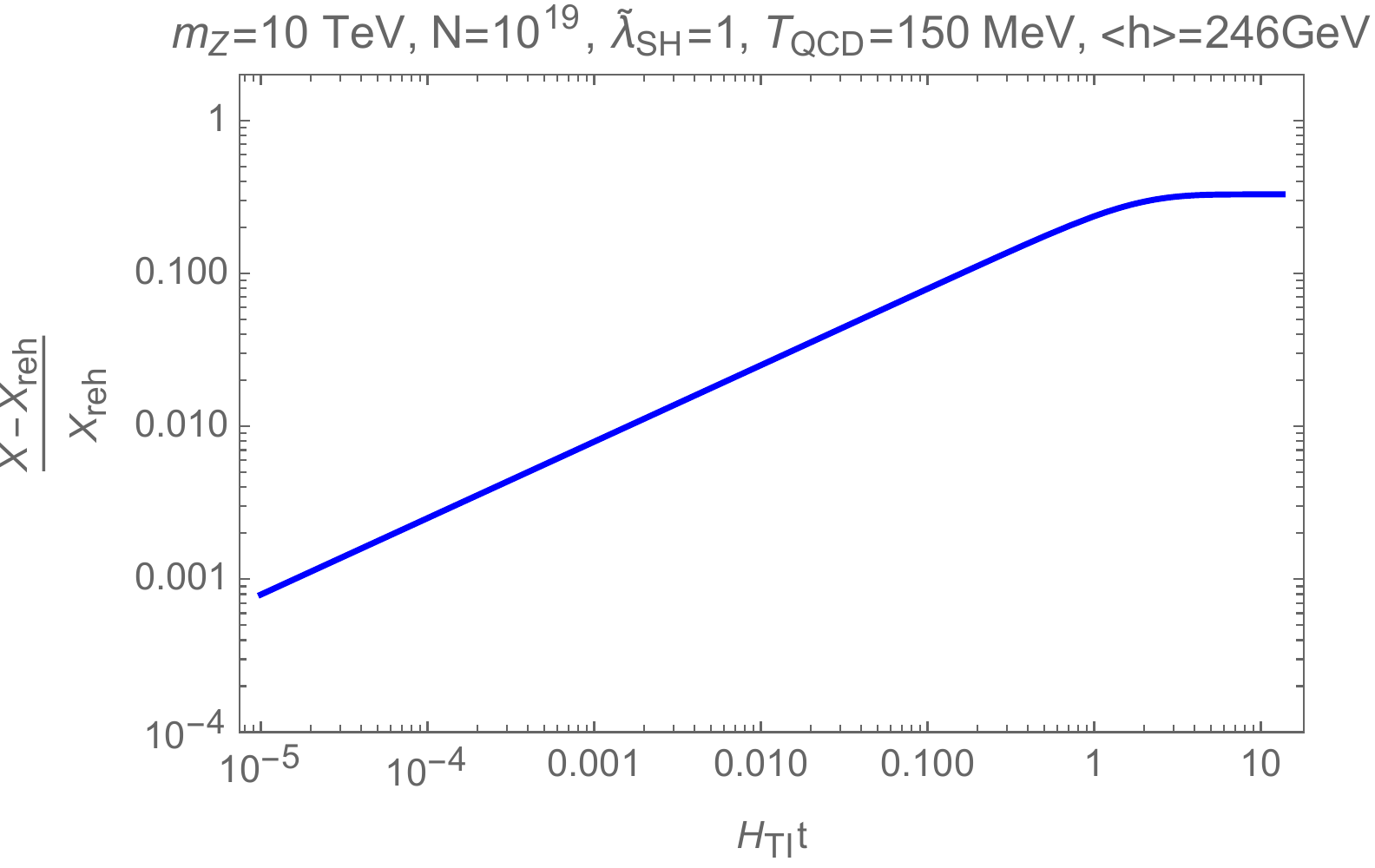}
\includegraphics[width=7cm]{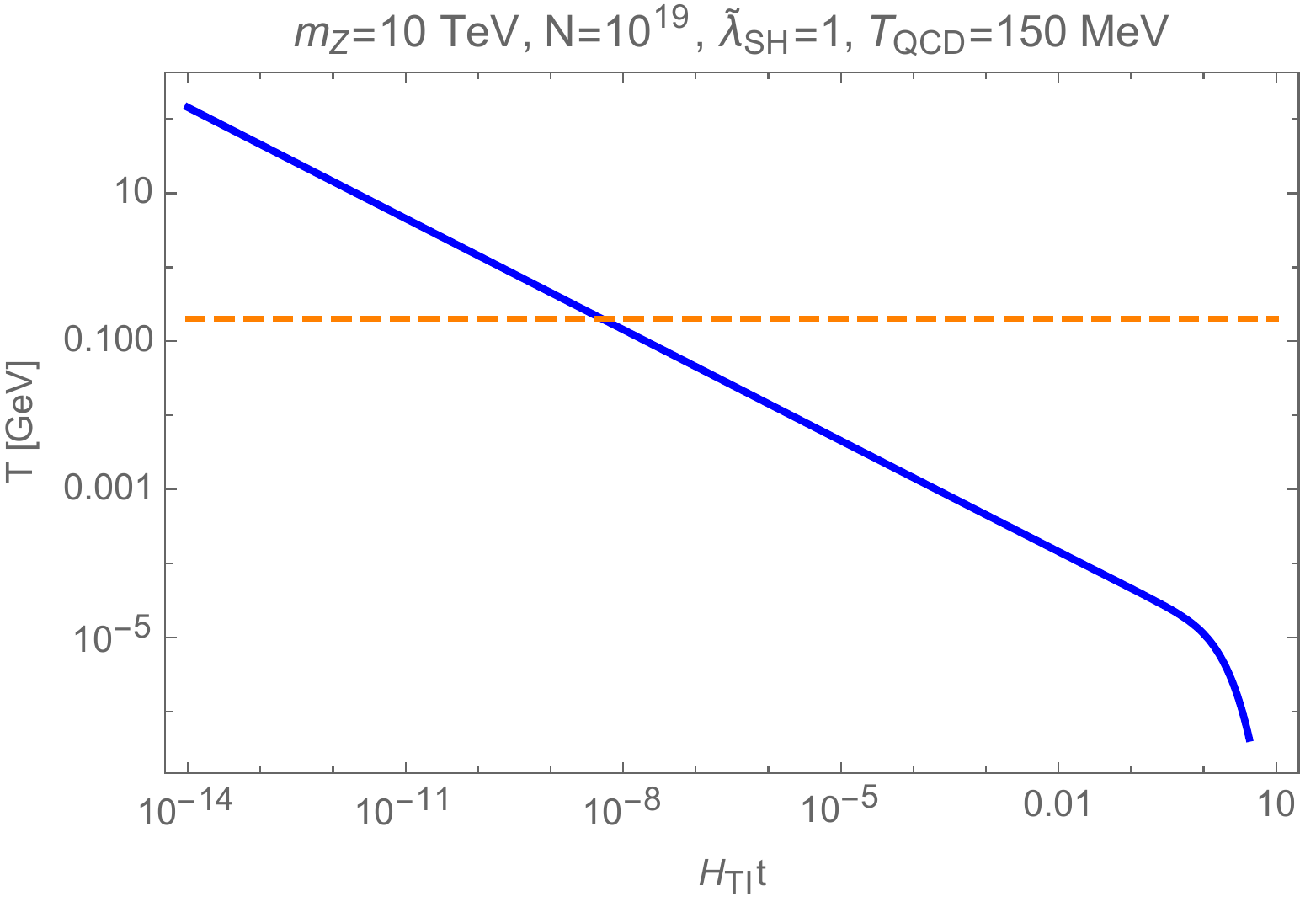}
\includegraphics[width=7cm]{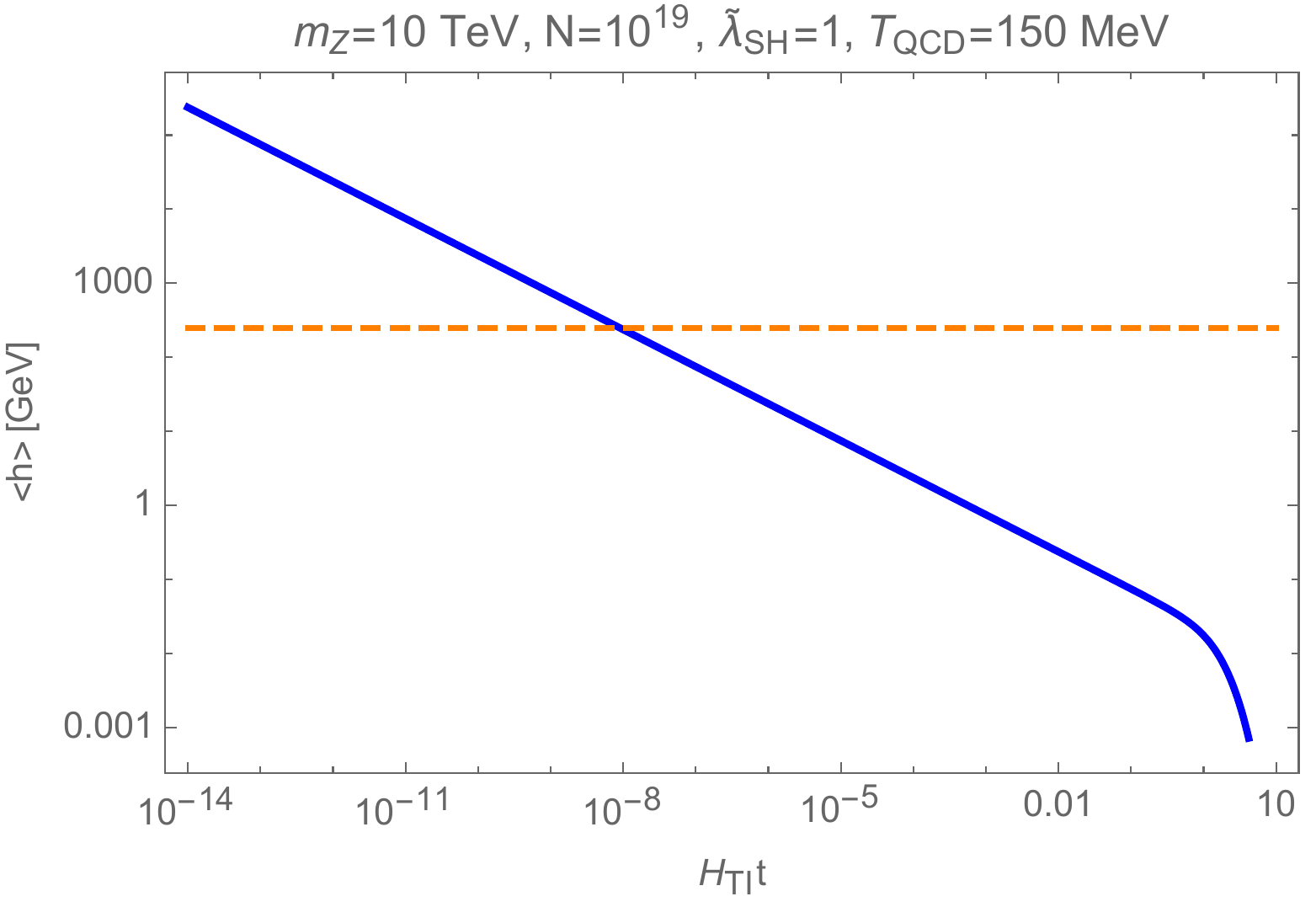}
\caption{Time evolutions of the ratio $X=\rho_S/\rho_{\rm SM}$ (top-left), temperature of the SM sector $T$ (top-right)  and the Higgs vev (down) in the massless case. }
\label{fig_massless}
\end{center}
\end{figure}

Then, the ratio $X=\rho_{\rm SM}^{}/\rho_S^{}$ obeys 
\aln{
 \frac{dX}{da}&= \frac{1}{aH}\frac{dX}{dt} =
 \frac{{\epsilon} \tilde{\lambda}_{SH}^2}{128\pi N^{9/4}}\left(\frac{30}{\pi^2}\right)^{5/4}\left(\frac{\zeta(3)}{\pi^2}\right)^{2}\frac{\rho_S^{1/4}}{aH}
\nn
&=c\times \begin{cases} 1  & \text{for $a\leq 1$}
  \\ 
            a^{-2}       & \text{for $a\geq 1$}
\end{cases}~,
\label{X equation relativistic}
}
where
\aln{
c&=\frac{{\epsilon}  \tilde{\lambda}_{SH}^2}{128\pi N^{9/4}}\left(\frac{30}{\pi^2}\right)^{5/4}\left(\frac{\zeta(3)}{\pi^2}\right)^{2}\frac{V_{\rm TI}^{1/4}}{H_{\rm TI}^{}}%=\frac{3^{1/2}\tilde{\lambda}_{SH}^2}{128\pi N^{9/4}}\left(\frac{30}{\pi^2}\right)^{5/4}\left(\frac{\zeta(3)}{\pi^2}\right)^{2}\frac{M_{pl}^{}}{V_{\rm TI}^{1/4}}~,
%\nn&
=\frac{3^{3/2}5^{5/4}\zeta(3)^2\tilde{\lambda}_{SH}^2}{2^4\pi^7 N^{9/4}}\frac{M_{pl}^{}}{m_{Z'}^{}}~. 
}
Eq.~(\ref{X equation relativistic}) can be solved as 
\aln{X-X_{\rm TI}^{}=c\times \begin{cases} a-1  & \text{for $a\leq 1$}
  \\ 
            1-a^{-1}       & \text{for $a\geq 1$}
\end{cases}~,
\label{solution of X in massless}
}
from which we obtain
\aln{
X_{\rm TI} \simeq X_{\rm reh}^{} + c \quad (\text{for $a_{\rm reh} \ll 1$})
}
and
\aln{
X\simeq X_{\rm TI}^{}+c \simeq  X_{\rm reh}^{} + 2c\quad (\text{for $a\gg 1$})~,
\label{Xeq}
}
where $X_{\rm reh}^{}$ comes from the SM energy density generated during the reheating after the primordial inflation. 
In the top-left panel of Fig.~\ref{fig_massless}, we show the time evolution of $(X-X_{\rm reh}^{})/X_{\rm reh}^{}$. 
Note that, compared to the massive case Eq.~(\ref{X solution in massive case}), the ratio $X$ keeps increasing and approaches a constant value $X_{\rm reh}^{}+2c$ because both of $S$ and SM particles behave as radiation and their energy densities dilute with $a^{-4}$. 
On the other hand, if  $X_{\rm reh}$ is larger than $2c$, 
\aln{
X_{\rm reh} > 2c \label{sufficient}
}
$X$ is almost constant from the onset of the $O(N)$-radiation dominated phase,
\aln{
X \simeq X_{\rm reh} ~. \label{Xconstant}
}
We will see  that this situation is necessary for the axion-CMB scenario in the massless $O(N)$ case 
because otherwise 
the CMB fluctuations transferred from the axions are diluted after QCD phase transition. 
%  
%%%%%%%%%%%%%%%%%%%%%%%%%%%
%%%%%%%%%%%%%%%%%%%%%%%%%%% 

\vspace{5mm}
%%%%%%%%%%%%%%%%%%%%%%%%%%%%%%%%%%%%%
\noindent$\bullet$ {\bf Various quantities at $T_{\rm QCD}$
}\\
The required value of $N$ is determined in a similar  way as the massive case. 
By using Eq.~(\ref{vev in massless case}), the condition for the Higgs vev at $T=T_{\rm QCD}^{}$ becomes
\aln{
X|_{T=T_{\rm QCD}^{}}^{}=\frac{\rho_{\rm SM}^{}}{\rho_S^{}}\bigg|_{T=T_{\rm QCD}^{}}^{}=\frac{g_{\rm SM}^{}}{N}\left(\frac{T_{\rm QCD}^{}}{\tilde{T}}\right)^4%=\frac{\rho_{\rm SM}^{}}{\rho_S^{}}
=\frac{g_{\rm SM}^{}}{N}\left(\frac{\tilde{\lambda}_{SH}^{}}{24 \lambda_H^{}}\right)^2\left(\frac{T_{\rm QCD}^{}}{\langle h\rangle}\right)^4~,
\label{X and h}
}
%where we used Eq.~(\ref{vev in massless case}). 
%
Let $\tilde{T}_{\rm QCD}$ denote the temperature $\tilde{T}$ of $S$ sector at $T=T_{\rm QCD}$. 
The  scale factor at this moment is then given by
\aln{
a_{\rm QCD}^{}:&=a|_{T=T_{\rm QCD}^{}}^{}=\frac{\tilde{T}_{\rm TI}^{}}{\tilde{T}_{\rm QCD}^{}} \nn
&\sim 1.4\times 10^{-4}\times \tilde{\lambda}_{SH}^{1/2}\left(\frac{0.1}{\lambda_H^{}}\right)^{1/2}\left(\frac{10^{19}}{N}\right)^{1/4}
\left(\frac{m_{Z'}^{}}{10~{\rm TeV}}\right)\left(\frac{246~{\rm GeV}}{\langle h\rangle}\right)~,
\label{scale factor at TQCD}
}
where we used Eqs.~(\ref{temperature in massless case}) and (\ref{vev in massless case}). 
Due to $H_{\rm TI}^{}t_{\rm QCD}^{}=a^2_{\rm QCD}\sim 2\times 10^{-8} \ll H_{\rm TI}^{}t_{\rm TI}=1$, 
it is much before the beginning of the thermal inflation.
This means that, if Eq.~(\ref{sufficient}) is not satisfied, 
the perturbations of $\rho_{\rm SM}$ generated at the QCD phase transition is diluted due to
further productions of the SM particles from the $O(N)$ scalar after the transition.
Hence, for realizing the axion-CMB scenario in explained in Section \ref{sec:review},  
we simply impose the condition Eq.~(\ref{sufficient}). 
Then  $X$ is almost constant as in Eq.~(\ref{Xconstant}) and it must be given by 
\aln{
X_{\rm reh} =\frac{g_{\rm SM}^{}}{N}\left(\frac{\tilde{\lambda}_{SH}^{}}{24 \lambda_H^{}}\right)^2\left(\frac{T_{\rm QCD}^{}}{\langle h\rangle}\right)^4 
}
to realize Eq.~(\ref{X and h}). 
Combined with Eq.~(\ref{sufficient}), it reads
\aln{
N > 7.9 \times 10^{17} \paren{\frac{\epsilon }{0.1}}^{4/5}  \paren{\frac{100}{g_{\rm SM}}}^{4/5} \paren{\frac{10~{\rm TeV}}{m_{Z'}}}^{4/5} \paren{\frac{\lambda_H}{0.1}}^{8/5} \paren{\frac{150~{\rm MeV}}{T_{\rm QCD}}}^{16/5} \paren{\frac{\langle h \rangle}{246 {\rm GeV}}}^{16/5} ~. \label{N in massless case}
}
As mentioned in Introduction, such a large value of $N$ is still allowed phenomenologically because collider observables are typically functions of $N\lambda_{SH}^{2}=\tilde{\lambda}_{SH}^{2}/N$ or $N\lambda_{SH}^{3}=\tilde{\lambda}_{SH}^{3}/N^2$. 
See Refs.~\cite{Meade:2018saz,Glioti:2018roy} for more details. 

\

%%%%%%%%%%%%%%%%%%%%%%%%%%%%%%%%
\noindent$\bullet$ {\bf Dilution factor by thermal inflation}\\
First, $\rho_S^{}$ is given by $V_{\rm TI}^{}$ at $\tilde{T}=\tilde{T}_{\rm TI}^{}$ by definition.  
The Higgs vev $\langle h\rangle$ decreases with temperature and finally approaches $v_{\rm QCD}$; %determined by the QCD chiral condensates in Eq.~(\ref{vQCD});
\aln{
\langle h\rangle^2=\begin{cases}\langle h\rangle^2|_{T^{}=T_{\rm QCD}}^{}\left(a_{\rm QCD}^{}/a\right)^2 & T\sim T_{\rm QCD}^{}
\\
v_{\rm QCD}^2 & T^{}\ll T_{\rm QCD}^{}
\end{cases}. 
\label{Higgs vev after TQCD}
}
We need to know the temperature $\tilde{T}_{\rm end}^{}$ at which the thermal inflation ends. 
%
%In presence of the $S$ sector, 
In the massless case, the quadratic term of the $B$-$L$ scalar $\phi$ of Eq.~(\ref{quadratic term in phi}) is modified to be
\aln{
&  \frac{T^2}{2} 
g_{B-L}^2\phi^2+
%\frac{\tilde{\lambda}_{S\phi}^{}}{4N}\langle S^2\rangle 
\frac{\tilde{\lambda}_{S\phi}^{}}{24}\tilde{T}^2\phi^2
-g_{B-L}^2\left(\frac{m_H^{}}{m_{Z'}}\right)^2\langle h\rangle^2 \phi^2~,  
\label{phi-potential in massless}
}
where the second term is the thermal mass correction by $S$ and this dominants over the first term as long as $\tilde{\lambda}_{S\phi}^{}>12g_{B-L}^2(T/\tilde{T})^2$. 
Then, the end of the thermal inflation is determined by the last  two terms of Eq.~(\ref{phi-potential in massless}) with $\langle h\rangle =v_{\rm QCD}^{}$~\footnote{As discussed below, $m_{Z'}^{}\sim 10~$TeV is allowed in the massless case because we can easily earn the dilution of $S$ compared to the massive case. 
For such a small value of $m_{Z'}^{}$, the third term in Eq.~(\ref{phi-potential in massless}) is already bigger than the first term due to the largeness of the Higgs vev $\langle h\rangle \sim 100~$GeV at $T=T_{\rm QCD}^{}$. 
}; 
\aln{
\tilde{T}_{\rm end}^{}= \sqrt{\frac{24}{\tilde{\lambda}_{S\phi}^{}}}\times g_{B-L}^{}\left(\frac{m_H^{}}{m_{Z'}}\right)v_{\rm QCD}^{}~. 
}
As a result, the dilution factor during the thermal inflation is given by %from $\tilde{T}=\tilde{T}_{H}^{}$ until $\tilde{T}=\tilde{T}_{\rm end}^{}$ is calculated as
\aln{
 e^{-4\Delta N}%\times a_H^{4} 
=\left(\frac{\tilde{T}_{\rm end}^{}}{\tilde{T}_{TI}^{}}\right)^4=\frac{30g_{B-L}^4}{\pi^2 N}\left(\frac{24}{\tilde{\lambda}_{S\phi}^{}}\right)^2\left(\frac{m_H^{}}{m_{Z'}^{}}\right)^4\frac{v_{\rm QCD}^{4}}{V_{\rm TI}}~,
\label{dilution in massless case}
}
where we have used the relation (\ref{massless density}) at $\tilde{T}=\tilde{T}_\tx{TI}$. 
For given values of $g_{B-L}^{}$ and $\tilde{\lambda}_{S\phi}^{}$, the above dilution factor becomes tiny if $m_{Z^\prime}$ is heavy.  
On the other hand, as we will see below,  $m_{Z^\prime}$ must be typically lighter than $10$~  TeV for sufficient dilution of axions.

\vspace{5mm}
%%%%%%%%%%%%%%%%%%%%%%%%%%%%
\noindent$\bullet$ {\bf Dilution of axions} \\
We now estimate $r_A^{}$ in the massless case.  
As well as the massive case, the oscillating temperature is determined by $m_{A0}^{}=3H$; 
\aln{
\tilde{T}_{\rm osc}^{}=\left(\frac{10}{\pi^2 N}\right)^{1/4}(m_{A0}^{}M_{pl}^{})^{1/2}%\left(\frac{T_{\rm QCD}^{}}{T}\right)^{2b}
=2.2\times 10^{-3}~{\rm GeV}\left(\frac{10^{19}}{N}\right)^{1/4}\left(\frac{m_{A0}^{}}{6\times 10^{-6}~{\rm eV}}\right)^{1/2}%\left(\frac{T_{\rm QCD}^{}}{T}\right)^{2b}
~, 
\label{Tosc in massless}
}
%which is consistent with $\tilde{T}_{\rm osc}^{}<\tilde{T}_H^{}$ i.e. $T_{\rm osc}^{}<T_{\rm QCD}^{}$. 
%
which is much smaller than $\tilde{T}_{\rm TI}^{}$. 
Thus, the axion field does not evolve until the beginning of thermal inflation and we can again use the same result of $r_A^{}$, %presented in Ref.~\cite{Iso:2020pzv} with different e-folding number:
\aln{
r_A^{} \sim \paren{\frac{2 \tan \paren{\bar{\theta}_{\rm ini}/2}}{0.3}}^2 \paren{\frac{f_A}{10^{12} {\rm GeV}} }^{1.16} \exp \paren{-2 \eta \times \Delta N}~, 
\label{r_A with thermal inflation} 
}
where $\eta$ is the same as Eq.~(\ref{axion slow-roll parameter}) and 
\aln{
%\eta :=\frac{1}{3} \paren{\frac{m_{A0}}{H_{\rm TI}}}^2=8.6\times 10^{-6}\times \left(\frac{10^{12}~{\rm GeV}}{f_A^{}}\right)^2\left(\frac{10~{\rm TeV}}{m_{Z'}^{}}\right)^4~,\quad 
\Delta N=\log \left(\tilde{T}_{\rm TI}^{}/\tilde{T}_{\rm end}^{}\right)~ . 
% \label{axion slow-roll parameter}
}
From the observational constraint for $R \sim 0.01$ in Fig.~\ref{fig:NG} and $r_A$ in Eq.~(\ref{r and R}), $r_A^{}$ must be as tiny as $10^{-4}$. 
We will thus plot the excluded region of $r_A^{}<8.2\times 10^{-3}R$ in the $\tilde{\lambda}_{SH}^{}$ - $m_{Z}'$ plane for a given value of $f_A^{}$.  

\

%%%%%%%%%%%%%%%%%%%%%%%%%%%%%%%%%%%%%%%%%
\noindent$\bullet$ {\bf Dilution of initial abundance of $S$}\\
Let us now evaluate the relic abundance of $S$ in the massless case.
Instead of Eq.~(\ref{rhos now}), the relic abundance of {massless} components of $S$ is estimated as 
\aln{
\frac{\rho_S^{}}{\rho_{\gamma}^{}}\bigg |_{\rm today}^{}
&=\frac{\rho_S^{}}{\rho_{\gamma}^{}}\bigg|_{T=T_R^{}}^{}
=\left(\frac{g_R^{} \pi^2}{30}\frac{T_{R}^4}{V_{\rm TI}}\right)^{4/3}\times e^{-4\Delta N}%\times a_{H}^4
\times \frac{V_{\rm TI}^{}}{\rho_\gamma^{}|_{T=T_{R}^{}}}
\nn
&=\frac{N-1}{2}\left(\frac{g_R \pi^2}{30}\frac{T_{R}^4}{V_{\rm TI}}\right)^{4/3}%\left(\frac{\lambda_H^{}}{b\tilde{\lambda}_{SH}^{}}\right)^2e^{-4\Delta N}\times 
\left(\frac{\tilde{T}_{\rm end}^{}}{T_{R}^{}}\right)^4
\nn
&\sim \frac{3.4\times 10^{-7}}{\tilde{\lambda}_{S\phi}^{2}}\left(\frac{N}{%9\times 10^{15} 
10^{19}}\right)\left(\frac{g_R^{}}{100}\right)^{4/3}\left(\frac{g_{B-L}^{}}{0.1}\right)^4\left(\frac{v_{\rm QCD}^{}}{150~{\rm MeV}}\right)^4
\left(\frac{10~{\rm TeV}}{m_{Z'}^{}}\right)^{28/3}\left(\frac{T_R^{}}{m_H^{}}\right)^{4/3}~.
\label{relativistic abundance of S}
}
The above relic abundance contributes to the present energy density as dark radiation. 
Extra contribution to the number of relativistic species is defined by  
\aln{
\rho_{\rm rad}^{}:=N_{\rm eff}^{}\frac{7}{8}\left(\frac{4}{11}\right)^{4/3}\rho_\gamma^{}~. 
}
The current bound by Planck 2018 \cite{Aghanim:2018eyx} is
\aln{N_{\rm eff}^{}=2.99^{+0.34}_{-0.33}\quad (95\%\ \rm {CL})~.
\label{bound on Neff} 
}
In the standard cosmology, $N_{\rm eff}^{}=3.046$ by neutrinos. 
Thus, Eq.~(\ref{relativistic abundance of S}) corresponds to
\aln{\Delta N_{\rm eff}^{}=\frac{8}{7}\left(\frac{11}{4}\right)^{4/3}\rho_S^{}/\rho_{\gamma}^{}|_{\rm today}^{}=1.5\times 10^{-6}\times \cdots~.  
}

On the other hand, the massive component of the $O(N)$ scalars %acquires mass $m_S = \sqrt{\lambda_{SH}} \langle h \rangle = \sqrt{\tilde{\lambda}_{SH} /N} \times 246$GeV and it 
becomes non-relativistic when the temperature $T$ is of the same order as $m_S^{}=(\tilde{\lambda}_{SH}^{}/N)^{1/2}\  v$. 
With $\rho_S|_{\rm today}$ being the energy density of the $N-1$ massless components, 
the energy density of the massive component is evaluated as 
\aln{
r_S^{}&=\frac{\rho^{\rm Massive}_S}{\rho_{\rm DM}}=\frac{1}{N-1}\frac{\rho_S^{}}{\rho_\gamma^{}}\bigg|_{\rm today}^{}\  (\rho_{\gamma}^{}/\rho_{\rm DM}^{})|_{\rm today}^{}\  \frac{m_S^{}}{T_0^{}}
\nn
&\sim 10^{-20}\times \frac{\tilde{\lambda}_{SH}^{1/2}}{\tilde{\lambda}_{S\phi}^{2}}\left(\frac{10^{19}}{N}\right)^{1/2}\left(\frac{g_R^{}}{100}\right)^{4/3}\left(\frac{g_{B-L}^{}}{0.01}\right)^4\left(\frac{v_{\rm QCD}^{}}{150~{\rm MeV}}\right)^4
\left(\frac{10~{\rm TeV}}{m_{Z'}^{}}\right)^{28/3}\left(\frac{T_R^{}}{m_H^{}}\right)^{4/3}~,
\label{abundance of massive component}
}
which is tiny and we can safely neglect the isocurvature of the $O(N)$ scalar.

\

%
%
%%%%%%%%%%%%%%%%%%%%%%%%%%%%%%%%%%%%%%%%%
\noindent$\bullet$ {\bf Productions of $S$ after thermal inflation}\\
We also have to consider the production of $S$ after the secondary reheating. 
Here we consider a case of low reheating temperature, $T_R^{}\leq m_H^{}$,  and focus on $hh\rightarrow SS$. 
In the following, we set $a=1$ at $T=T_R^{}$ i.e. $T=T_R^{}a^{-1}$.  
The Boltzmann equation is
\aln{\dot{\rho}_S^{}+4H\rho_S^{}=\frac{N\lambda_{SH}^2}{16\pi m_H^{}}n_H^2=\frac{\tilde{\lambda}_{SH}^2}{16\pi Nm_H^{}}\left(\frac{m_H^{}T}{2\pi}\right)^3e^{-2m_H^{}/T}~,
}
from which the ratio $Y:=\rho_S^{}/\rho_{\rm SM}^{}$ obeys 
\aln{\frac{dY}{da}&=\frac{\tilde{\lambda}_{SH}^2}{16\pi Nm_H^{}}\left(\frac{m_H^{}T}{2\pi}\right)^3\frac{e^{-2m_H^{}/T}}{aH\rho_{\rm SM}^{}}
%\nn
%&=\frac{3^{1/2}M_{pl}^{}\tilde{\lambda}_{SH}^2}{16\pi Nm_H^{}}\left(\frac{m_H^{}T}{2\pi}\right)^3\frac{e^{-2m_H^{}/T}}{a\rho_{\rm SM}^{3/2}}
\nn
&=d (m_H^{}/T_R^{})  (m_H^{}a/T_R^{})^2e^{-2(m_H^{}/T_R^{})a}~,
\label{equation of Y}
} 
where we defined
\aln{
d:&=\frac{3^{1/2}\tilde{\lambda}_{SH}^2}{2^7\pi^4 N}\left(\frac{30}{\pi^2g_{\rm SM}^{}}\right)^{3/2}\frac{M_{pl}^{}}{m_H^{}%T_R^{}
\label{defofd}
}
\nn
&=1.4\times 10^{-9}\times \tilde{\lambda}_{SH}^2\left(\frac{100}{g_{\rm SM}^{}}\right)^{3/2}\left(\frac{10^{19}}{N}\right)~. %\left(\frac{m_H^{}}{T_R^{}}\right). 
}
The integration of Eq.~(\ref{equation of Y}) gives 
\aln{Y\sim d/4\quad \text{for $a\gg T_R^{}/m_H^{}$}~, 
\label{production Y}
}
which contributes to the dark radiation as well as the initial abundance of $S$.   
Thus, if $N$ is sufficiently large with $\tilde{\lambda}_{SH}$ fixed, we can safely avoid the over-production of $S$. 
As for the massive component, it is easy to see that its production is also negligible by the same calculation as 
Eq.~(\ref{abundance of massive component}).  
In~\ref{ffSS}, we also discuss the productions via $f\bar{f}\rightarrow SS$, which is found to be subdominant compared to $hh\rightarrow SS$.  

\begin{figure}[t!]
\begin{center}
\includegraphics[width=6.9cm]{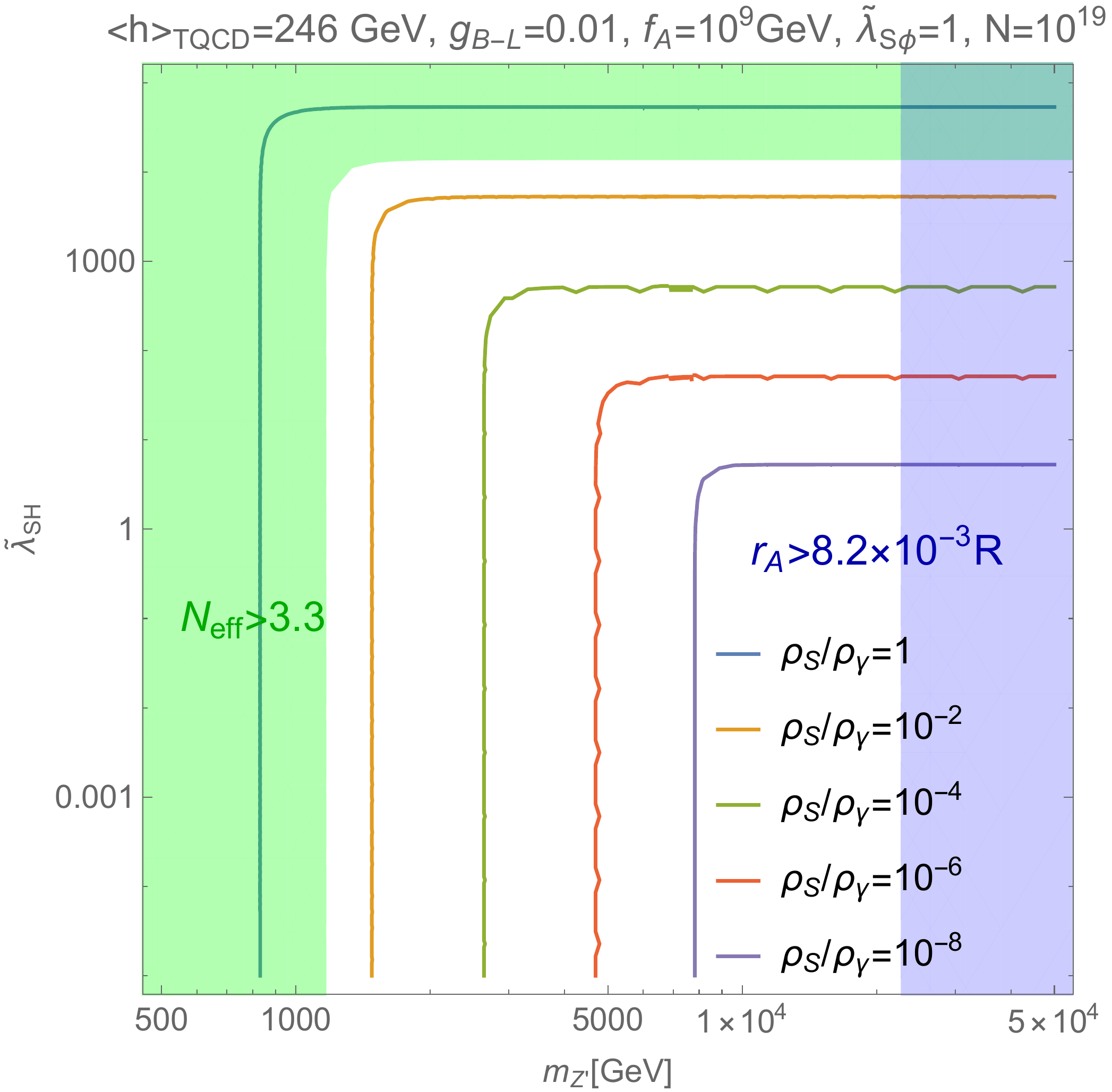}
\includegraphics[width=7cm]{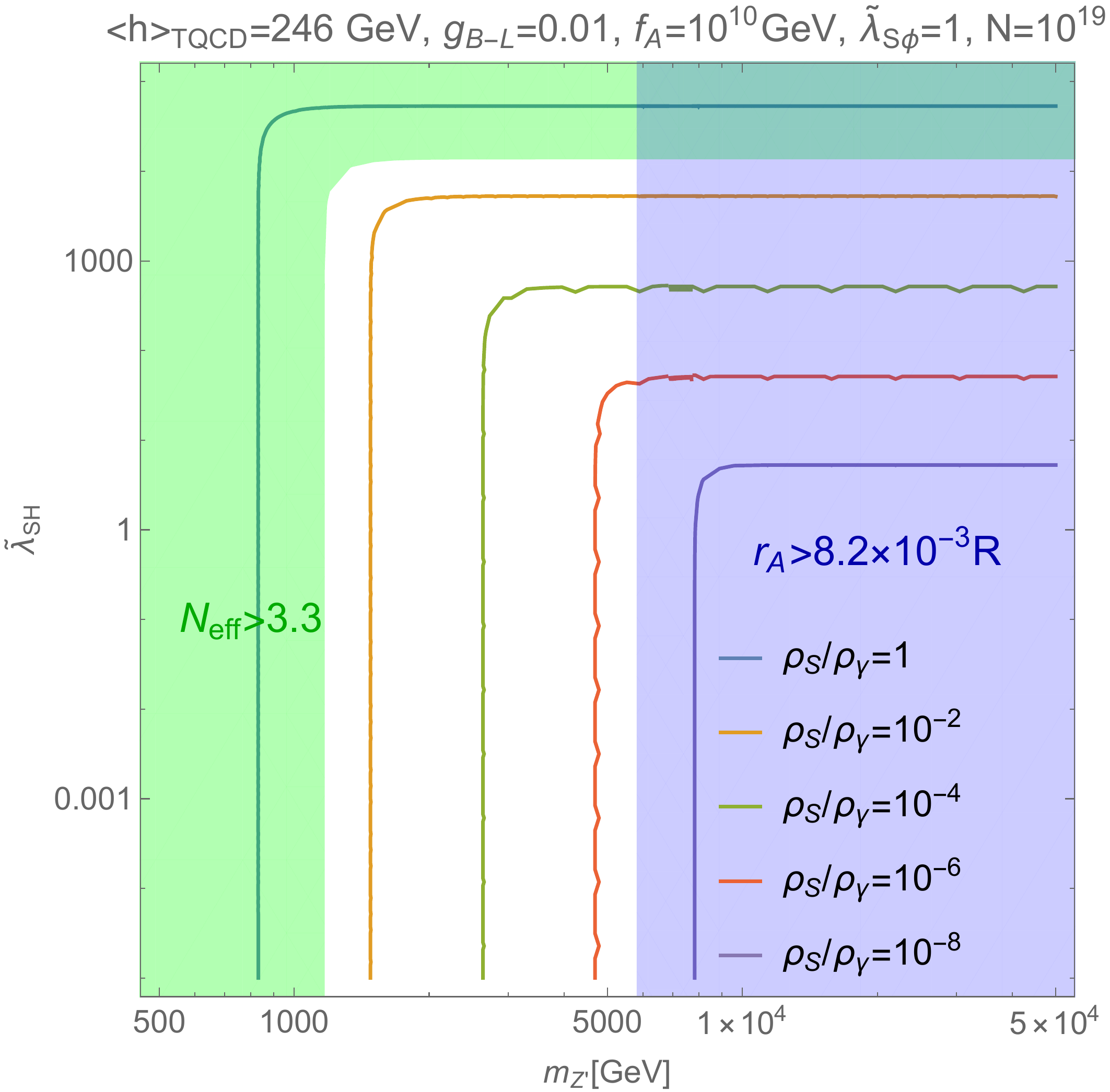}
\caption{
Allowed parameter regions in the case of massless $S$. 
The left (right) panel corresponds to $f_{A}=10^9~(10^{10})~{\rm GeV}$.  
}
\label{fig:region_massless}
\end{center}
\end{figure}

In Fig.~\ref{fig:region_massless}, we show the allowed parameter region on the ($\tilde{\lambda}_{SH}$, $m_{Z^\prime}$)  plane. 
The left (right) panel corresponds to  $f_{A}=10^9~(10^{10})~{\rm GeV}$.  
% 
%%%%%%%%%%%%%%%%%%%%%%%%%%%
%%%%%%%%%%%%%%%%%%%%%%%%%%%    
In the figure, the green regions are excluded by the constraint on the dark radiation Eq.~(\ref{bound on Neff}) while the blue regions are excluded by the isocurvature constraint of the CMB observations, $r_A^{} < 8.2\times 10^{-3}R$.     
As a result, one can see that $m_{Z'}^{}$ is constrained to be
$1~{\rm TeV}\lesssim m_{Z^\prime}\lesssim 22~{\rm TeV}~(6~{\rm TeV})$ when $f_A^{}=10^9~(10^{10})~$GeV.

%________________________________________
\section{Conclusions}
In the paper, we have investigated a possibility of the axion-CMB scenario 
 in  particle physics models with {\it classical conformality (CC).}
In such CC models, the early universe often experiences supercooled era of the EW symmetry breaking 
until the temperature of the universe cools down to QCD scale $T_{\rm QCD}$, 
and the thermal inflation naturally occurs at low temperature. 
On the other hand, the axion-CMB scenario in which the CMB fluctuations are assumed to 
be generated from
the primordial axion fluctuations requires that the axion abundance must be sufficiently
diluted after the axion potential is generated at $T_{\rm QCD}$. 
Thus, if the thermal inflation in the CC models occurs below $T_{\rm QCD}^{}$, 
CC models can be candidates for particle physics models realizing the axion-CMB scenario.

In this paper, we have particularly studied the CC $B$-$L$ model with an additional extension of $O(N)$  scalars $S$. 
Such an extension is necessary in order to generate sufficiently large amplitude of axion potential at QCD temperature. 
Namely, the Higgs vev in the supercooled era of EW symmetry 
must be generated by a negative scalar coupling with $S$ and become as large as $10^2$ GeV. 
This requirement is fulfilled by considering decoupled evolutions of the SM sector and $O(N)$ sector where the temperature of each sector is completely different. 

We have investigated various observational constraints of CMB observations and the present abundance of 
axions and the additional particles $S$, and shown that, if $S$ is massless,
there is a parameter region in which all the constraints are satisfied.

%%%%%%%%%%%%%%%%%%%% ACKNOWLEDGMENTS %%%%%%%%%%%%%%%%%%%%
\section*{Acknowledgements} 
%We would like to thank XX for useful discussions.   
%
%The work of KK is supported by the Grant-in-Aid for JSPS Research Fellow, Grant Number 17J03848.  
The work of SI is supported in part by the Grant-in-Aid for Scientific research, No. 18H03708, No. 16H06490.

%%%%%%%%%%%%%%%%%%%%%%%%%%%Appendix%%%%%%%%%%%%%%%%%%%%%%%%%%%%%%%%%%%%%%%%
\appendix 
\def\thesection{Appendix \Alph{section}}

%________________________________________________
\section{Boltzmann equations}\label{app:Boltzmann}
In this appendix, we summarize the basic calculations of Boltzmann equations \cite{Gondolo:1990dk,Ala-Mattinen:2019mpa}. 

\

%_______________________________________
\noindent $\bullet$ {\bf Thermal average of cross sections}\\
The thermal average of cross section by the Maxwell-Boltzmann distribution is given by \cite{Gondolo:1990dk}
\aln{\langle \sigma v\rangle=\frac{1}{T^2xK_2^{}(x)^2}\int_1^\infty dy(4m^2\sigma)(y-1)&\sqrt{y}K_1^{}(2x\sqrt{y}),\ x=m/T,\ y=s/(4m^2)~,
%\\
%&(\text{for Maxwell-Boltzman distribution})\nonumber
}
where 
\aln{
K_\nu^{}(z)=\frac{\pi^{1/2}(z/2)^\nu}{\Gamma(\nu+1/2)}\int_1^\infty dte^{-zt} (t^2-1)^{\nu-1/2} 
}
is the modified bessel function of the second kind. 
The cross sections of $SS\leftrightarrow HH$ are  
\aln{
\sigma_{SS\rightarrow HH}^{}(s)=\frac{\lambda_{SH}^2}{16\pi s}\sqrt{\frac{s-4m_H^2}{s-4m_S^2}}%=\frac{\lambda_{SH}^2}{64\pi m_S^2(1+\epsilon)}\sqrt{\frac{\epsilon+4(m_S^2-m_h^2)}{\epsilon}}
~,\quad \sigma_{HH\rightarrow SS}^{}(s)=\frac{\lambda_{SH}^2}{16\pi s}\sqrt{\frac{s-4m_S^2}{s-4m_H^2}}%=\frac{\lambda_{SH}^2}{64\pi m_S^2(1+\epsilon)}\sqrt{\frac{\epsilon+4(m_S^2-m_h^2)}{\epsilon}}
~,
}
from which we obtain
\aln{
&\langle \sigma_{SS\rightarrow HH} v\rangle=\frac{\lambda_{SH}^2}{16\pi \tilde{T}^2x_S^{}K_2^{}(x_S^{})^2}\int_{(m_H^{}/m_S^{})^2}^\infty dy y^{-1/2}\sqrt{(y-1)(y-(m_H^{}/m_S^{})^2)}K_1^{}(2x_S^{}\sqrt{y})~,\label{thermal average SSHH}
\\
&\langle \sigma_{HH\rightarrow SS} v\rangle=\frac{\lambda_{SH}^2}{16\pi T^2x_{H}^{}K_2^{}(x_{H}^{})^2}\int_{(m_S^{}/m_H^{})^2}^\infty dy y^{-1/2}\sqrt{(y-1)(y-(m_S^{}/m_h^{})^2)}K_1^{}(2x_H^{}\sqrt{y})~, 
\label{thermal average HHSS}
}
where $x_S^{}=m_S^{}/\tilde{T}$, $x_{H}^{}=m_H^{}/T^{}$~. 
Our main focus is Eq.~(\ref{thermal average SSHH}) because we assume that $S$ is the dominant component during the period between the primordial preheating and the thermal inflation.  
In the massive $S$ case, we saw that $m_H^{}/m_S^{}$  is always negligible as shown in Eq.~(\ref{mass ratio}). 
%At around the beginning of thermal inflation, we have 
%\aln{\left(\frac{m_H^{}}{m_S^{}}\right)^2\sim \frac{\tilde{\lambda}_{SH}^{}}{128\pi^2N}\left(\frac{m_{Z^\prime}^{}}{m_S^{}}\right)^4\ll1~, 
%} 
%where we haves used Eqs.~(\ref{ns and v}). %(\ref{number density at TTI})
%
Thus, we can neglect the higher order terms of $(m_H^{}/m_S^{})^2$ in the RHS in Eq.~(\ref{thermal average SSHH}).   
Within this approximation, the integration becomes  
\aln{
&\int_1^\infty dy(y-1)^{1/2}K_1^{}(2x_S^{}\sqrt{y})=K_1^{}(x_S^{})^2/2x_S^{}~,
%\\
%&\int_1^\infty dyy^{-1/2}(y-1)^{1/2}K_1^{}(2x_{H}^{}\sqrt{y})=\frac{\pi}{4x_H^2}e^{-2x_H^{}},
}
which leads to
\aln{
\langle \sigma_{SS\rightarrow HH}^{} v\rangle=\frac{\lambda_{SH}^2}{32\pi \tilde{T}^2x_S^2}\left(\frac{K_1^{}(x_S^{})}{K_2^{}(x_S^{})}\right)^2~.
%\quad \langle \sigma_{HH\rightarrow SS}^{} v\rangle=\frac{\lambda_{SH}^2T_{\rm SH}^{}e^{-2x_H^{}}}{64m_{H}^3 K_2^{}(x_H^{})^2}~. 
}
When $T\gg m$, i.e. $x\ll 1$, the bessel functions are expanded as  
\aln{
\left(\frac{K_1^{}(x)}{K_2^{}(x)}\right)^2=\frac{x^2}{4}+{\cal O}(x^3)~,%\quad \frac{1}{K_2^{}(x)^2}=\frac{x^4}{4}+{\cal O}(x^5)~,
}
from which we have 
\aln{
&\langle \sigma_{SS\rightarrow HH}^{}v\rangle\sim\frac{\lambda_{SH}^2}{128\pi \tilde{T}^2} \quad \text{for $\tilde{T}\gg m_S^{}$}~, 
%\\
%&\langle \sigma_{HH\rightarrow SS}^{}v\rangle\sim \frac{}{}
\label{average at high temperature}~.
}
On the other hand, in the non-relativistic limit $\tilde{T}\ll m_S^{}$, i.e. $x_S^{}\ll 1$, the bessel functions are expanded as 
\aln{
\left(\frac{K_1^{}(x)}{K_2^{}(x)}\right)^2=1-\frac{3}{x}+{\cal O}(x^{-2})~,
}
from which we have   
\aln{
\langle \sigma_{SS\rightarrow HH}^{} v\rangle \sim \frac{\lambda_{SH}^2}{16\pi m_S^2}\left(1-\frac{3\tilde{T}^{}}{m_S^{}}\right)\quad \text{for $\tilde{T}^{}\ll m_S^{}$}~.
%\frac{A}{\sqrt{\pi}m^2}\left(\frac{T}{m}\right)^{1/2}. 
}
%This is natural because the typical velocity can be estimated as  
%\aln{T\sim \frac{m}{2}v^2\quad \rightarrow\quad v\sim \left(\frac{T}{m}\right)^{1/2}. 
%}  

\

%________________________________________________
\noindent $\bullet$ {\bf Boltzmann equations}\\
In general, the Boltzmann equation for particle species $1$ is given by
\aln{ \frac{\partial n_1^{}}{\partial t}+3Hn_1^{}=\int \frac{d^3\mathbf{p}}{(2\pi)^3E(p)}\hat{C}[f_1^{}]
}
where  the collision term is
\aln{
\int C[f_{1}]\frac{d^{3}p}{(2\pi)^{3}}=-\sum_{\text{dof}}\int&\frac{d^{3}p_{1}}{(2\pi)^{3}2E_{1}}\frac{d^{3}p_{2}}{(2\pi)^{3}2E_{2}}\frac{d^{3}p_{3}}{(2\pi)^{3}2E_{3}}\frac{d^{3}p_{4}}{(2\pi)^{3}2E_{4}}\nonumber\\
&\times \{f_{1}f_{2}(1\pm f_{3})(1\pm f_{4})|{\cal{M}}_{12\rightarrow34}|^{2}-f_{3}f_{4}(1\pm f_{1})(1\pm f_{2})|{\cal{M}}_{34\rightarrow12}|^{2}\}\nonumber\\
&\times(2\pi)^{4}\delta(p_{1}+p_{2}-p_{3}-p_{4}).
}
In our model, $1,2$ corresponds to $S_i^{}\ (i=1,2,\cdots,N)$, and $3,4$ corresponds to the SM Higgs, and vice versa.  
When the system is not experiencing BEC and Fermi degeneracy, it is good to approximate $(1\pm f_l^{})$ by 1.  
Then, assuming thermal (kinetic) equilibrium in each particles,  we have 
\aln{
\dot{n}_i^{}+3Hn_i^{}=-\langle \sigma_{SS\rightarrow HH}^{} v\rangle n_i^2+\langle \sigma_{HH\rightarrow SS}^{} v\rangle n_{H}^2~. 
} 
where $\langle\sigma_{SS(HH)\rightarrow HH(SS)} v\rangle$ is the thermal average Eqs.~(\ref{thermal average SSHH})(\ref{thermal average HHSS}).   
%

%%%%%%%%%%%%%%%%%%%%%%%%%%%%%%%%%%%%
\section{Reheating temperature after thermal inflation}\label{app:T_R}
In this appendix, 
we give a rough estimation of the reheating temperature $T_R$ after the thermal inflation.
%in Eq. (\ref{reheating temperature}). 
It is defined as the temperature of the SM radiation at the moment 
when the coherent oscillation of the flaton field $\phi$ is disintegrated into radiation.
We assume that the reheating is completed  either by $\phi$'s decay into two SM Higgs particles or 
by $\phi$'s scattering with the Higgs in the SM thermal bath.
Reheating  by $\phi$ decay into SM particles via the scalar mixing is neglected for simplicity.

We restrict ourselves to the case where the $B-L$ gauge boson $Z^\prime$ is much heavier than the right-handed neutrino $N$ and 
the coefficient in Eq.~(\ref{CWpotential}) is dominantly given by $Z^\prime$. 
Then, the mass of $\phi$ field at the true  vacuum is obtained as $m_\phi \sim g_{B\hyphen L} m_{Z^\prime}$.
If $m_\phi \gtrsim m_h $, or equivalently,
\aln{
g_{B\hyphen L} \gtrsim m_h /m_{Z^\prime} ~, \label{Cond.Decay}
}
$\phi$ can decay into two Higgs particles with the rate
\aln{
\Gamma_{\rm d} \sim \frac{\paren{\lambda_{\phi H} v_{\phi} }^2 }{m_{\phi}} \sim g_{B\hyphen L} \frac{m_h^4}{m_{Z^\prime}^3} ~.
}
Equating this with the Hubble expansion rate $H \sim T^2 /M_{pl}$, 
we get 
\aln{
T_R^{\rm (d1)} = \paren{g_{B\hyphen L} m_{Z^\prime} M_{pl}}^{1/2} \paren{m_h / m_{Z^\prime}}^2.
\label{reheatingTdecay1}
}
The derivation is valid as far as the thermal mass of the Higgs is below $m_\phi$.
The condition  $m_h(T_R^{(d1)}) \sim T_R^{(\rm d1)} \lesssim m_\phi$ is rewritten as
\aln{
g_{B\hyphen L} \lesssim x:= \frac{m_h^4 M_{pl}}{m_{Z^\prime}^5}.
\label{Cond.Blocked}
}
If the condition is not satisfied, the decay process is kinematically forbidden 
until  the temperature is reduced to  $T = m_\phi \sim g_{B\hyphen L} m_{Z^\prime}$. 
In this  case, the reheating temperature is given by $T_R^{(\rm d2)} = m_\phi \sim g_{B\hyphen L} m_{Z^\prime}$.
To summarize,  the reheating temperature by decay is 
\aln{
T_{R}^{(\rm d)} \sim {\rm min} \br{T_R^{\rm (d1)}, m_\phi }
%m_\phi %g_{B\hyphen L} m_{Z^\prime} 
%\times {\rm min} \br{1 ,  \paren{\frac{M_{pl}}{g_{B\hyphen L} m_{Z^\prime} }}^{1/2} \paren{\frac{m_h}{m_{Z^\prime}}}^2} 
\leq m_\phi .
\label{reheatingTDecay}
}
Note that the scalar field $\phi$ acquires mass through the Coleman Weinberg mechanism, and  
$m_\phi$ is generally small. In the CC $B$-$L$ model, in particular, it cannot be much heavier than the SM Higgs; $m_\phi \le m_H^{}$.  
Thus the reheating temperature by the decay process is also expected to be not so high compared to $m_H^{}$.

On the other hand, the scattering process is not thermally blocked.
Once the SM thermal bath is generated, 
the coherent oscillation is disintegrated into $\phi$ particles due to scattering with the Higgs in the bath.
%In the nonrelativistic region $T \lesssim m_\phi$, since $\Gamma_{\rm s} \propto T^3/m_\phi^2$, 
The interaction rate in the relativistic region $T \gtrsim m_\phi$ is given by
\aln{
\Gamma_{\rm s} &\sim  \lambda_{\phi H}^2  T   %& {\rm for}~~T \gtrsim m_{\phi}  
%\Gamma_{\rm s} &\sim  \lambda_{\phi H}^2 \times \left\{ \begin{matrix} T & {\rm for}~~T \gtrsim m_{\phi}  \\ T^3/m_{\phi} &  {\rm for}~~T \lesssim m_{\phi} \end{matrix} \right. \nn
\sim   g_{B\hyphen L}^4 \paren{\frac{m_h}{m_{Z^\prime}}}^4  T ,
}
whereas $\Gamma_{\rm s} \sim  \lambda_{\phi H}^2 T^3/m_{\phi}$ for $T \lesssim m_{\phi}$. 
By comparing it with the Hubble, %$\Gamma_{\rm s} \sim H$, 
the reheating temperature by the scattering process is given by \footnote{The $\phi$ particles  produced in the scattering process  have yet to decay or pair-annihilate into the SM particles. Therefore, this $T_{R}^{(\rm s)}$ should be regarded as a  maximum 
value of the reheating temperature.}
\aln{
T_{R}^{(\rm s)} \sim  g_{B\hyphen L}^4  M_{pl} \paren{\frac{m_h}{m_{Z^\prime}}}^4 
\label{reheatingTscat}
}
as far as  $T_R^{(\rm s)} \gtrsim  m_\phi$ is satisfied, i.e. 
\aln{
g_{B\hyphen L}^3 \gtrsim \frac{m_{Z^\prime}^5}{m_h^4 M_{pl}} =1/x ~.
\label{Cond.Scattering}
}
If it is not satisfied, we have $T_{R}^{(\rm s)} = 0$. 
For example, if $m_{Z^\prime}=10$~TeV and $g_{B-L}=0.01$, $T_R^{(s)}\sim 600$ GeV.

The reheating temperature is given  by 
$ T_R =  {\rm max} \br{ T_R^{(\rm d)} ,T_{R}^{(\rm s)}  } $.
Comparing  Eq.~(\ref{reheatingTDecay})    and  Eq.~(\ref{reheatingTscat}), 
we have $T_R= T_R^{(\rm s)}$  when  the condition (\ref{Cond.Scattering}) %$T_R^{(\rm s)} \gtrsim  m_\phi$
is satisfied.
%The condition  $T_R^{(\rm s)} \gtrsim  m_\phi$ is  rewritten as 
%\aln{
%g_{B\hyphen L}^3 \gtrsim \frac{m_{Z^\prime}^5}{m_h^4 M_{pl}} =1/x ~. 
%\label{Cond.Scattering}
%}
 Provided $g_{B\hyphen L} < 1$, 
when the condition (\ref{Cond.Scattering}) is violated, %i.e., $T_R^{(\rm s)} <  m_\phi$, 
the inequality $x > 1/g_{\rm B\hyphen L}^3> g_{\rm B\hyphen L}$
follows and Eq.~(\ref{Cond.Blocked}) is always satisfied. 
Thus as far as  Eq.~(\ref{Cond.Decay}) is satisfied, 
\aln{
T_{R} \sim   T_R^{(\rm d1) }  < m_\phi^{}
~~~ {\rm for} ~~~ g_{B\hyphen L} \gtrsim \frac{m_h }{m_{Z^\prime} } ~.
}
If Eq.~(\ref{Cond.Decay}) is not satisfied, the decay process is forbidden, and the reheating is only possible because of the scattering process.
Then, the condition (\ref{Cond.Scattering}) must be satisfied so that $T_{R} = T_{R}^{(\rm s)}$ is non-zero.

%%%%%%%%%%%%%%%%%%%%%%%%%%%%%%%%%%%%%%%%%%%%%%%%
\section{$S$ production during reheating}\label{production during reheating}
In the body of the paper, we have considered
the production of the $O(N)$ scalar particles after the reheating is discussed.
Here, we will consider the production during the reheating of thermal inflation and show that
it does not surpass the production after the reheating. 

Neglecting the Boltzmann suppression,
the Boltzmann equation for the energy density is given as
\aln{
\dot{\rho}_S + 4 H \rho_S \sim N \lambda_{SH}^2 T \rho_{\rm SM} ~,
}
where $\rho_{\rm SM}\sim T^4$ and $H \simeq \sqrt{\rho_\phi /3 M_{pl}^2} = H_R a^{-3/2}$ with $H_R = \sqrt{\rho_\phi /3 M_{pl}^2}|_{T=T_R}$.
$\rho_{\rm SM}$ first grows by its production from coherent oscillation of $\phi$,  
and then it is expected to decrease with a scaling coefficient $x$ 
as  $T=T_{R} a^{-x}$. After the reheating, the coefficient approaches $x=1.$ 
Since the energy density of coherent oscillation behaves as $a^{-3}$ and $\rho_\phi \sim \rho_{\rm SM}$
 at $T_R$,  $x$ is expected to be 
$x<3/4$ so that $\rho_{\rm SM} < \rho_{\phi}$ before $T_R$.

Then, the Boltzmann equation becomes
\aln{
\frac{d (\rho_S a^4)}{da} \sim  \frac{\wt{\lambda}_{SH}^2}{N} \frac{T_R^5}{H_R} a^{\frac{9}{2}- 5 x}
}
and is solved as
\aln{
\rho_S \sim \rho_{S,{\rm end}} \paren{\frac{a_{\rm end}}{a}}^4 + \frac{d ~ T_{R}^4 a^{-4} }{11/2 - 5 x}   \paren{ a^{\frac{11}{2}- 5 x}  - a^{\frac{11}{2}- 5 x}_{\rm end}} 
}
with $d$ defined in Eq.~(\ref{defofd}).
%Remember that it is required that the preexisting $O(N)$ particles are well diluted during the thermal inflation.
Since $\frac{11}{2}- 5 x > 0$, when  the reheating process ends at $a=1$, 
the energy density of produced $S$ is given by
\aln{
\left. \rho_S \right|_{T=T_R} \sim d ~ T_{R}^4
}
because of $a_{\rm end} \ll 1$.
Thus the  $S$ particles produced early in the reheating period are diluted and the ones produced around $a=1$ is dominant,
which justifies the analysis in the body of the paper. 

%%%%%%%%%%%%%%%%%%%%%%%%%%%%%%%%%%%%
%________________________________________
\section{$f\bar{f}\rightarrow SS$}\label{ffSS}
In this appendix,
 we consider  production of $S$ after the secondary reheating via the process $f\bar{f}\rightarrow SS$ where $f$ is a SM fermion.  
The cross section is
\aln{\sigma_{f\bar{f}\rightarrow SS}^{}(s)=\frac{y_f^2v^2\lambda_{SH}^2}{64\pi}\frac{1}{(s-m_H^2)^2+m_H^2\Gamma_H^2}\frac{s-4m_f^2}{s}~,
}
where $\Gamma_H^{}=4~{\rm MeV}$ is the total decay width of the SM Higgs. 
The thermal average is now given by
\aln{\langle \sigma_{f\bar{f}\rightarrow SS}^{} v\rangle=&\frac{v^2\lambda_{SH}^2}{64\pi T^2xK_2^{}(x)^2\times 4m_f^2}\int_1^\infty dy\frac{(y-1)^2y^{-1/2}}{(y-m_H^2/(4m_f^2))^2+m_H^2\Gamma^2/(4m_f^2)^2}K_1^{}(2x\sqrt{y})~,
\\
 &x=m_f^{}/T,\ y=s/(4m_f^2)~. 
}
%%%%%%%%%%%%%%%%%%%%%%%%%%%
\begin{figure}[t!]
\begin{center}
\includegraphics[width=9cm]{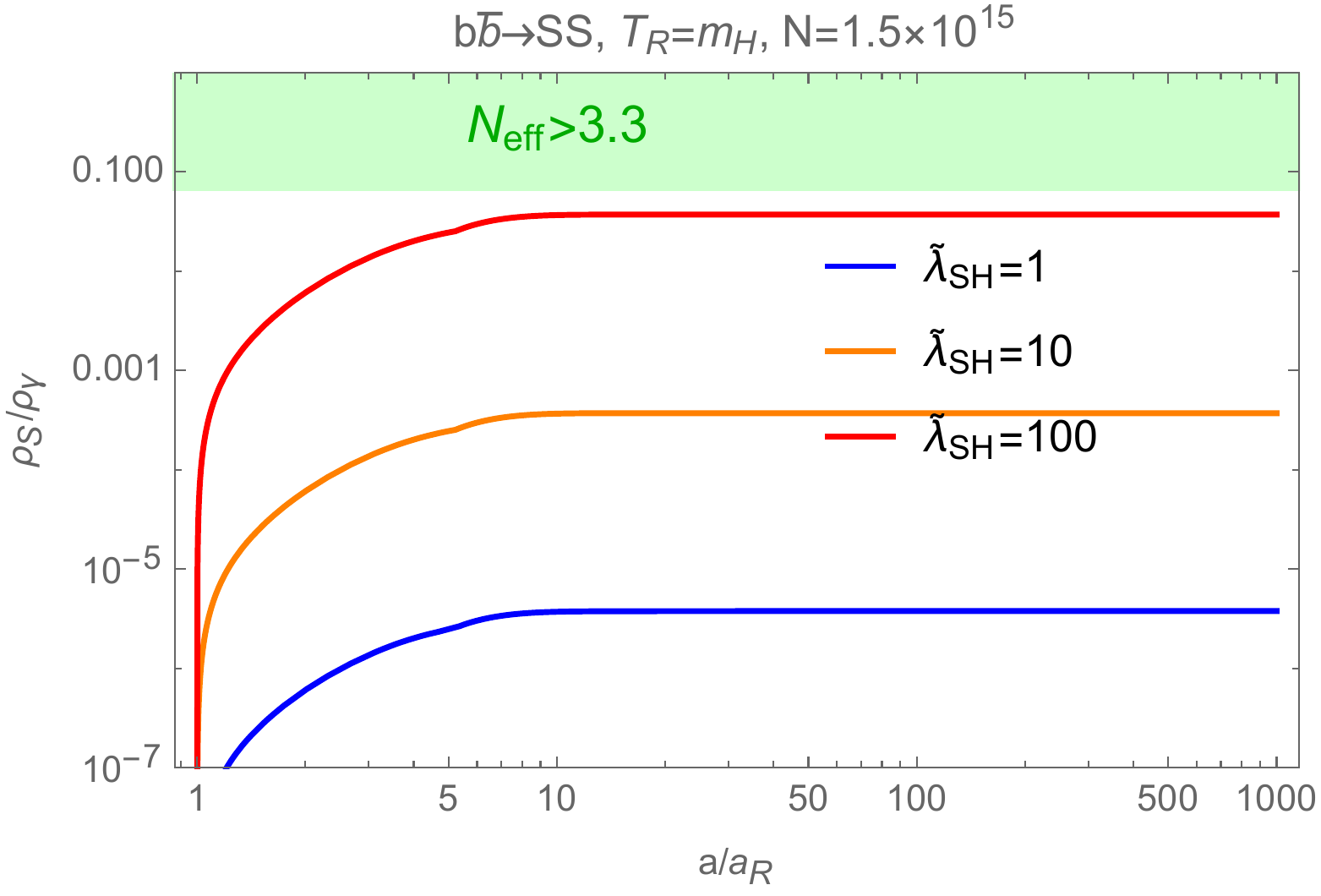}
\caption{Production of $S$ via $b\bar{b}$ after the reheating of thermal inflation.  
}
\label{fig:bbSS}
\end{center}
\end{figure}
%%%%%%%%%%%%%%%%%%%%%%%%%%%    
The Boltzmann equation is
\aln{\dot{\rho}_S^{}+4H\rho_S^{}=NT\langle \sigma_{f\bar{f}\rightarrow SS}^{} v\rangle n_f^{}(T)^2\quad ,\quad \rho_S^{}|_{T=T_R^{}}=0~,
\label{eq:ffSS}
}
where
\aln{
n_f^{}(T)=\int \frac{d3p}{(2\pi)^3}\frac{1}{e^{\beta E(p)}+1}\quad,\quad E(p)=\sqrt{p^2+m_f^2}~.
}
In Fig.~\ref{fig:bbSS}, we show the numerical calculations of Eq.~(\ref{eq:ffSS}) where the different colors correspond to the different values of $\tilde{\lambda}_{SH}^{}$. 

\bibliography{Bibliography}
\bibliographystyle{utphys}

\end{document}